\documentclass[a4paper,fleqn]{cas-sc}
\usepackage{url}
\usepackage{setspace}
\usepackage{geometry}
\usepackage{bm}
\usepackage{tabularray}
\usepackage[numbers]{natbib}
\usepackage{makecell}
\def\tsc#1{\csdef{#1}{\textsc{\lowercase{#1}}\xspace}}
\tsc{WGM}
\tsc{QE}
\tsc{EP}
\tsc{PMS}
\tsc{BEC}
\tsc{DE}

\usepackage[normalem]{ulem}
\newcommand{\stkout}[1]{\ifmmode\text{\sout{\ensuremath{#1}}}\else\sout{#1}\fi}

\makeatletter
\newcommand*{\rom}[1]{\expandafter\@slowromancap\romannumeral #1@}
\makeatother

\usepackage[mathscr]{euscript}
\usepackage{lineno}

\begin{document}
\let\WriteBookmarks\relax
\def\floatpagepagefraction{1}
\def\textpagefraction{.001}

\shorttitle{Robust Topology Optimization against First-Order Worst-Case Perturbations}

\shortauthors{J Wang et~al.}

\title[mode = title]{Sensitivity Hot Spot Penalization: A Robust Topology Optimization Framework against First-Order Worst-Case Perturbations}
 
\author[1]{Junpeng Wang}[
                        orcid=0000-0002-4607-844X] 
\cormark[1]
\ead{junwa@dtu.dk}

\credit{Conceptualization, Investigation, Methodology, Software, Writing - Original Draft}

\affiliation[1]{organization={Technical University of Denmark},
    country={Denmark}}

\author[1]{Niels Aage}[
                        orcid=0000-0002-3042-0036] 
\ead{naage@dtu.dk}
\credit{Conceptualization, Methodology, Writing - Review \& Editing, Supervision}

\author[1]{Ole Sigmund}[
                        orcid=0000-0003-0344-7249] 
\ead{olsi@dtu.dk}
\credit{Conceptualization, Methodology, Writing - Review \& Editing, Supervision, Funding acquisition}

\cortext[cor1]{Corresponding author}

\begin{abstract}
Deterministic topology optimization can efficiently generate high-performance structural designs, but it does not explicitly control localized structural fragility induced by manufacturing variations and geometric uncertainties. Such fragility often appears as localized stress concentrations or hinge-like deformation mechanisms. Conventional robust topology optimization can suppress these features, but typically requires multiple design realizations and substantially increased computational cost. Recent work~\cite{sigmund2026} introduced sensitivity hot spot penalization (SHoSP), in which the nominal objective is augmented by a smooth maximum of its sensitivities, and demonstrated that it can suppress localized fragile features at low additional cost, as well as indirectly generating more even stress distributions.

This paper establishes a general connection between SHoSP and the first-order worst-case robust approximation under a material-mass perturbation budget through a Taylor expansion and H{\"o}lder's duality, identifying the original penalty weight as a dimensionless norm-bounded material-mass perturbation budget relative to the area (2D) or volume (3D) of the design domain.
This explains the observed suppression of sensitivity hot spots, hinge localization, and stress concentrations resulting from reducing the maximum first-order degradation under budgeted perturbations. The interpretation is investigated for compliance minimization and compliant mechanism design, the two canonical mechanical settings considered in the original SHoSP study. Robustness is assessed by applying a relaxed worst-case density perturbation derived from the associated uncertainty model and comparing deterministic and SHoSP designs under equivalent perturbation budgets. 
The stress-related effects associated with SHoSP, as observed on structured meshes, are cross-validated by stress-oriented topology optimization and body-fitted finite element analyses.
Extensions to porous infill optimization, multiple load cases, and large-scale 3D examples further demonstrate the applicability and superior scalability of the SHoSP framework.
\end{abstract}


\maketitle
\onehalfspacing

\section{Introduction} \label{sec:Intro}
Topology optimization can generate highly efficient structural layouts. However, if optimized only for a nominal geometry and material distribution~\cite{bendsoe1989optimal, bendsoe2013topology}, small deviations introduced during manufacturing may cause a disproportionate loss of performance, particularly when the optimized response depends strongly on thin structural members, localized compliant regions, or other geometrically sensitive features~\cite{schevenels2011robust, wang2011projection}. This has motivated extensive research on robust topology optimization, in which manufacturing and geometric uncertainties are incorporated directly into the optimization formulation.

Existing approaches to topology optimization under uncertainty can be broadly distinguished by how uncertainty is represented. In probabilistic formulations, uncertain geometry, material properties, or loading conditions are modeled as random variables, random fields, or stochastic processes, and robustness is quantified through statistical moments, failure probabilities, or expected responses. For geometric manufacturing uncertainty, Schevenels et al.~\cite{schevenels2011robust} represented spatially varying errors in a density-based framework using a Karhunen--Loève expansion and evaluated the stochastic response through Monte Carlo sampling, whereas Chen and Chen~\cite{chen2011new} considered random boundary perturbations in a level-set formulation using numerical quadrature. To reduce the cost of uncertainty propagation, subsequent studies employed sparse-grid stochastic collocation~\cite{lazarov2012topologySMO}, non-intrusive polynomial-chaos expansions~\cite{keshavarzzadeh2017topology}, and stochastic level-set or nodal-perturbation descriptions of geometric variability~\cite{zhang2017robust,kim2022modeling}. Stochastic-gradient formulations have also been developed to update the design without fully resolving the stochastic response at each iteration~\cite{de2020topology, uihlein2026140}. The value of these detailed probabilistic descriptions is problem dependent: for compliance minimization and compliant mechanism design, spatially varying random errors may lead to designs similar to those obtained using simpler erosion--dilation models~\cite{schevenels2011robust}, whereas localized random perturbations can be substantially more important in dynamic and wave-propagation problems, where they may shift or suppress resonant responses~\cite{christiansen2015creating,elbek2025tailoring}.

By contrast, non-probabilistic set- or scenario-based formulations describe uncertainty through prescribed admissible realizations and optimize the structural response against representative or worst-case scenarios. The most prominent example in density-based topology optimization is the erosion--intermediate--dilation formulation, in which different projection thresholds generate geometries representing over- and under-machining, and the design is optimized with respect to the worst response among these realizations~\cite{sigmund2009manufacturing,wang2011projection}. Besides improving tolerance to manufacturing variations, this formulation provides explicit control of minimum feature sizes, suppresses localized hinges in compliant mechanism design, and promotes nearly discrete manufacturable layouts~\cite{qian2013topological,jansen2013similarities,lazarov2016length}. These advantages have made the three-field formulation a widely used reference for geometric robustness~\cite{da2021three}.
Compared with detailed stochastic descriptions, scenario-based formulations restrict the treatment of uncertainty to a small number of prescribed geometric realizations. Direct treatment, nevertheless, requires the structural response to be evaluated for each realization. This can be particularly relevant for compliant mechanisms in the large-scale 3D context, where erosion may weaken or disconnect slender hinges and transmission paths, causing the most adverse realization to become poorly conditioned and more difficult to solve with iterative solvers, for which the cost depends not only on the number of realizations but also on the numerical properties of each realization.

The computational expense associated with repeated uncertainty realizations has motivated perturbation-based approaches that approximate uncertain structural responses locally using sensitivity information~\cite{lazarov2012topologyIJNME}. Rather than explicitly resolving the response for every realization, these methods employ Taylor expansions about a nominal configuration to estimate the effect of uncertain parameters or geometric variations~\cite{guo2009confidence}. Within the erosion–dilation framework, Taylor expansions with respect to the projection threshold have been investigated to approximate the eroded and dilated responses from the intermediate realization, thereby reducing the cost of explicitly analyzing all three geometries~\cite{mommeyer2024taylor}. In probabilistic settings, a closely related approach is the first-order second-moment (FOSM) method. For robust compliance topology optimization, Kriegesmann and L{\"u}deker~\cite{kriegesmann2019robust} used a first-order expansion of the reduced objective together with the covariance of the uncertain parameters to approximate its statistical moments, and derived an efficient formulation requiring only one additional adjoint system to solve per optimization iteration. More recent developments generalized this idea to arbitrary response functions and non-intrusive principal-sensitivity evaluations~\cite{kruger2023efficient}, while node-wise geometric-uncertainty formulations investigated higher-order sensitivity aggregations that place greater emphasis on locally critical perturbations~\cite{karnath2026numerical}. These approaches demonstrate that derivative information can provide low-cost approximations of the effects of uncertainty without explicitly resolving a large number of uncertain realizations. The resulting robustness measure, however, remains determined by the adopted uncertainty description, perturbation parameterization, and, in probabilistic formulations, the assumed probability model, covariance structure, and selected risk measure.

Inspired by the observation that localized peaks in the objective sensitivity field indicate strong susceptibility to local design perturbations, Sigmund recently introduced sensitivity hot spot penalization (SHoSP)~\cite{sigmund2026}, in which the nominal objective is augmented by a smooth maximum of its area (volume)-normalized sensitivities with respect to the physical densities. SHoSP was shown to alleviate stress concentrations and suppress localized hinge-like mechanisms in compliant mechanism design, while requiring only two additional adjoint right-hand sides associated with the original stiffness matrix. It is worth noting that SHoSP shares several formal and computational features with FOSM-based approaches~\cite{kriegesmann2019robust}: both construct a low-cost robustness correction from first-order objective sensitivities and reuse the nominal equilibrium system in the associated sensitivity analysis. Their underlying viewpoints, however, are different. FOSM starts from a prescribed probabilistic model and quantifies global response variability through statistical moments, whereas SHoSP was introduced from the local observation that sharp sensitivity peaks identify structurally fragile regions and should therefore be directly suppressed.
Also, FOSM is formulated in terms of the mathematical design variable, whereas SHoSP is based on the physical variable and then propagated to the design variables through the chain rule.
In the original study~\cite{sigmund2026}, SHoSP was also compared to the conventional three-field robust formulation~\cite{wang2011projection} for compliant mechanism design under erosion/dilation perturbations, showing that SHoSP was substantially more robust than the deterministic design, although the three-field robust design performed somewhat better, especially under stronger erosion.
Despite the promising results of SHoSP~\cite{sigmund2026}, the robustness model implicitly represented by SHoSP was not theoretically explained: the associated uncertainty set was unspecified, the weighting parameter $\kappa$ was interpreted primarily as a fixed penalty coefficient, and the roles of the aggregation exponent and the area (volume)-normalization lacked a direct uncertainty-based explanation. Besides, the mechanical connection between sensitivity hot spot suppression and the observed reductions in stress concentrations and hinge localization was not formally clarified.

This work revisits SHoSP and establishes that it is the first-order worst-case approximation associated with a budgeted perturbation of the element-wise material mass. By combining a first-order Taylor expansion with H{\"o}lder's duality, the SHoSP term is recovered as the maximum first-order objective deterioration over a norm-bounded perturbation set~\cite{ben1998robust, guo2009confidence, thore2017general}. This identifies $\kappa$ as a dimensionless norm-bounded material-mass perturbation budget relative to the area or volume of the design domain, explains the normalized sensitivity as the objective derivative with respect to local material content, and relates the aggregation exponent to the spatial concentration permitted for the perturbation. The resulting interpretation is complemented by a mechanical analysis showing that sensitivity hot spots correspond to density-weighted stress-energy concentrations in compliance minimization and to localized state-adjoint interaction in compliant mechanism design. 
The theory is assessed using density perturbations constructed from the associated uncertainty model, with deterministic and SHoSP designs compared under equivalent perturbation budgets. The stress-related effects observed on structured meshes are cross-validated by stress-oriented topology optimization and body-fitted finite element analyses. Extensions to porous infill optimization, multiple load cases, and large-scale 3D examples further demonstrate the applicability and superior scalability of the framework.

The remainder of this paper is organized as follows. Section~\ref{sec:TO} introduces the density-based topology optimization setting used throughout the paper. In Sec.~\ref{sec:NewInterp}, we first revisit the original SHoSP formulation and then establish its mathematical connection to first-order worst-case robustness under a budgeted material-mass perturbation uncertainty set. Section~\ref{sec:stressSensHotSpots} clarifies the mechanical meaning of sensitivity hot spots, with emphasis on their connection to stress concentrations in compliance minimization and localized hinge mechanisms in compliant mechanism design. Section~\ref{sec:implScale} discusses implementation and scalability aspects of the method. Section~\ref{sec:results} presents numerical examples that examine the robustness interpretation, mechanical effects, and scalability of SHoSP. Finally, conclusions are drawn in Sec.~\ref{sec:conclusion}.


\section{Density-based topology optimization} \label{sec:TO}

In its original form, the mathematical program describing density-based topology optimization for isotropic and linear elasticity is given by
\begin{align}
    \displaystyle \min \limits_{\bm{x}} \quad & \Phi(\bm{U}(\bm{\rho})), \label{eqn:obj}\\
    \displaystyle \mathrm{s.t.} \quad & \bm{K}\bm{U}=\bm{F}, \label{eqn:FEA}\\
    & g(\bm{\rho}) \leq 0, \label{eqn:cons}\\
    & \bm{\rho}=\mathcal{P}(\mathcal{F}(\bm{x},R),\beta,\eta), \label{eqn:regul}\\
    & x_e \in [0,1], \:\: \forall e. \label{eqn:DV}
\end{align}
Here, $\Phi$ measures the structural performance to be optimized, while $g$ represents the optimization constraint(s). $\bm{x}$ denotes the design variable vector, $\bm{\rho}$ the physical density vector, $\bm{F}$ the nodal load vector, $\bm{U}$ the corresponding displacement vector, and $\bm{K}$ the global stiffness matrix. The physical densities ($\bm{\rho}$) are obtained by sequentially applying a filtering operation $\mathcal{F}$ and a projection operation $\mathcal{P}$ to the design variables ($\bm{x}$).

The material properties of the finite elements are interpolated using the \emph{Solid Isotropic Material with Penalization} (SIMP) scheme
\begin{equation}\label{eqn:SIMP}
    E_e(\rho_e) = E_{\min} + \rho_e^{q}(E_0-E_{\min})
\end{equation}
where $q$ is the penalization factor, typically chosen as $q=3$. $E_0$ denotes the Young's modulus of the solid material and $E_{\min}$ is a minimum stiffness introduced to avoid singular stiffness matrices. Throughout this work, $E_{\min}=10^{-6}E_0$.

To regularize the optimization problem and suppress mesh-dependent patterns, the Helmholtz-type PDE filter proposed by Lazarov and Sigmund~\cite{lazarov2011filters} is employed. The filtering operation is written as
\begin{equation}\label{eqn:PDEfilter}
    \hat{\bm{x}} =\mathcal{F}(\bm{x},R)= \bm{T}^{T}\left(l_0^2\bm{K}_{\mathcal{F}} + \bm{M}_{\mathcal{F}}^{vol} + l_s \bm{M}_{\mathcal{F}}^{surf}\right)^{-1}\bm{T}\bm{x}
\end{equation}
Here, $\hat{\bm{x}}$ denotes the filtered design variables, $l_0=R/(2\sqrt{3}h_e)$ is related to the filter radius $R$, and $h_e$ represents the element size. The matrices $\bm{K}_{\mathcal{F}}$ and $\bm{M}_{\mathcal{F}}^{vol}$ are assembled in a manner analogous to the stiffness and mass matrices in the finite element method, with each node carrying a single scalar degree of freedom (DOF). The matrix $\bm{T}$ maps between the element-wise and node-wise representations of the density-related vectors. The additional matrix $\bm{M}_{\mathcal{F}}^{surf}$ originates from the Robin boundary treatment proposed by Wallin et al.~\cite{wallin2020consistent}, which compensates for missing neighborhood information outside the design domain, thereby providing a more consistent filtering operation near boundaries. A simple exploration of $l_s$ and its effect can be found in Appendix~\ref{apdx:RobinBC}. Throughout this paper, we let $l_s=2l_0$, unless otherwise specified. 

The filtered densities ($\hat{\bm{x}}$) are subsequently projected to the physical densities ($\bm{\rho}$) through a smooth Heaviside projection
\begin{equation}\label{eqn:Heaviside}
    \rho_e(\hat{x}_e) = \mathcal{P}(\hat{x}_e,\beta,\eta) = \frac{\tanh(\beta\eta) + \tanh[\beta(\hat{x}_e-\eta)]}{\tanh(\beta\eta)+\tanh[\beta(1-\eta)]}
\end{equation}
where $\beta$ controls the sharpness of the projection and $\eta$ defines the threshold value. In this paper, $\eta = 0.5$ is adopted, and $\beta$ is increased through a continuation process. Specifically, $\beta$ is initialized as a small value not exceeding 1.0 and doubled every 50 optimization steps, and ends at a maximum value of $\beta_{\max}$. Accordingly, the initial value of $\beta$ is determined by $\beta=\frac{\beta_{\max}}{2^{\lfloor\log_2(\beta_{\max})\rfloor+1}}$. Unless otherwise specified, we let $\beta_{\max}=\frac{2R}{h_e}$. 

In its simplest form, the constraint function ($g$) is constructed as a global volume fraction constraint,
\begin{equation}\label{eqn:GVF}
    g(\bm{\rho}) = \frac{\sum_e \rho_e A_e}{V_0 \sum_e A_e}-1\leq 0
\end{equation}
where $A_e$ denotes the area (2D) or volume (3D) of element ($e$) and $V_0$ is the prescribed volume fraction.

We consider two representative optimization problems throughout this work, i.e., compliance minimization and compliant mechanism design. For compliance minimization,
\begin{equation}\label{eqn:Compliance}
    \Phi = \bm{F}^{T}\bm{U}
\end{equation}
where minimizing $\Phi$ maximizes the structural stiffness.

For compliant mechanism design, the output displacement is defined through a signed probing vector $\bm{L}$, and the objective function is written as
\begin{equation}\label{eqn:Mechanism}
    \Phi = -\bm{L}^{T}\bm{U}.
\end{equation}
Here, $\bm{L}$ defines the output quantity prescribed by the user. Its non-zero entry is assigned to the output DOF, and its sign is chosen according to the desired output direction. For multiple or distributed output DOFs, the non-zero entries distribute the output measure over the prescribed output region, such that $\bm{L}^{T}\bm{U}$ represents an averaged or integrated output displacement. To avoid a mesh-dependent output value, $\bm{L}$ is normalized as $\sum_i L_i = \pm 1$ and distributed over a fixed geometric length. The negative sign in Eq.~\ref{eqn:Mechanism} only converts the maximization of the prescribed output displacement into a minimization problem for optimization.

The sensitivity of the objective function with respect to the physical density field is computed by adjoint analysis. For compliant mechanism design, the element-wise sensitivity is given by
\begin{equation} \label{eqn:MechanismSensitivity}
    \frac{\mathrm{d} \Phi}{\mathrm{d} \rho_e} = \bm{\lambda}^{T} \frac{\partial \bm{K}}{\partial \rho_e} \bm{U}
\end{equation}
with
\begin{equation} \label{eqn:MechanismAdjoint}
    \bm{K}\bm{\lambda} = \bm{L},
\end{equation}
where $\bm{\lambda}$ is the adjoint vector. For compliance minimization, since it is a self-adjoint problem, i.e.,  $\bm{\lambda}=-\bm{U}$. The sensitivity expression, therefore simply reduces to
\begin{equation} \label{eqn:ComplianceSensitivity}
    \frac{\mathrm{d} \Phi}{\mathrm{d} \rho_e} = -\bm{U}^{T} \frac{\partial \bm{K}}{\partial \rho_e} \bm{U}
\end{equation}

Note that the sensitivities above are defined with respect to the physical densities ($\bm{\rho}$). Before being supplied to the optimizer, they are back-propagated through the projection and filtering operations using the chain rule, as documented in~\cite{andreassen2011efficient, wang2025efficient}.


\section{SHoSP as a first-order worst-case approximation} \label{sec:NewInterp}

\begin{figure}
    \centering
    \includegraphics[width=1.0\linewidth]{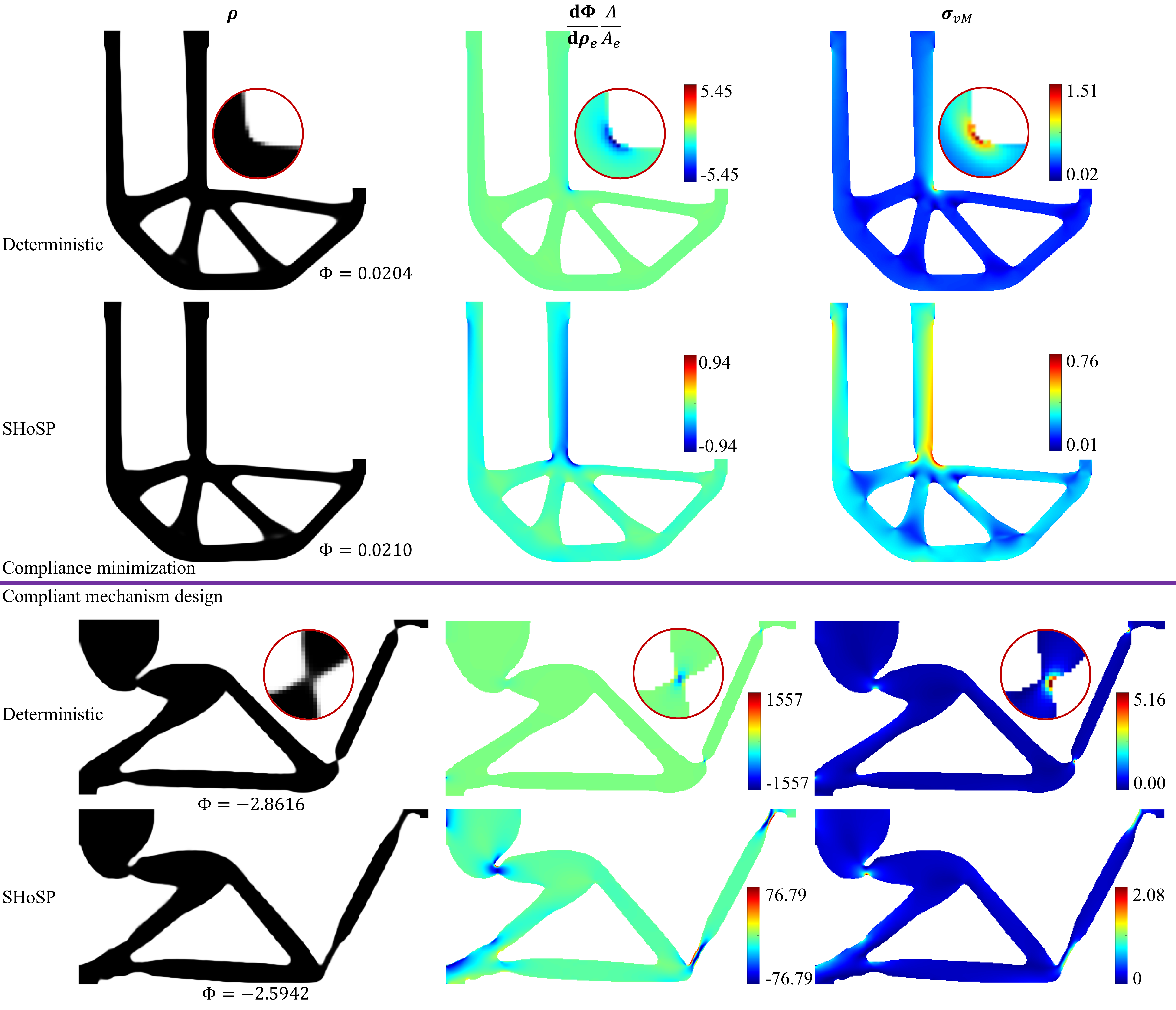}
    \caption{Representative effects of SHoSP in compliance minimization and compliant mechanism design. Deterministic and SHoSP designs are compared using the physical density $\bm{\rho}$, the area- or volume-normalized objective sensitivity $\frac{d \Phi}{d \rho_e}\frac{A}{A_e}$, and the von Mises stress field $\bm{\sigma}_{vM}$, where $A$ and $A_e$ denote the areas (volumes for 3D) of the design domain and element, respectively. SHoSP suppresses localized sensitivity hot spots and alleviates the associated stress concentration or hinge-like localization with only moderate changes in nominal performance. Refer to Fig.~\ref{fig:BenchmarkPDs} for the detailed problem descriptions.}
    \label{fig:Recap_SHoSP}
\end{figure}

Before diving into the first-order worst-case interpretation, we briefly revisit the main observation underlying SHoSP~\cite{sigmund2026}. Figure~\ref{fig:Recap_SHoSP} shows two representative examples from compliance minimization and compliant mechanism design. In the deterministic designs, localized fragile features coincide with sharp peaks in the sensitivity field: a stress concentration near the re-entrant corner in the compliance problem and a hinge-like deformation mechanism in the compliant mechanism. When the objective is augmented by the SHoSP term, these sensitivity peaks are strongly suppressed and the corresponding stress or deformation localizations are redistributed over large regions. These observations motivate the SHoSP formulation reviewed below and raise the central question addressed in this section: what robustness model is implicitly represented by penalizing the sensitivity hot spots?

\subsection{SHoSP} \label{subsec:SHoSP}
The sensitivity hot spot penalization method, i.e., SHoSP, recently suggested by Sigmund~\cite{sigmund2026} is formulated through the augmented objective
\begin{equation}\label{eqn:augObj}
\tilde{\Phi} = \Phi + \kappa \Psi
\end{equation}
where $\Psi$ denotes a measure of sensitivity hot spots, and $\kappa$ is a scaling factor controlling the penalization intensity. Since sensitivity hot spots are characterized by localized peaks in the sensitivity field, a natural choice is to define the hot spot measure in terms of the maximum absolute sensitivity of the physical density and approximate it using a smooth $\ell_p$-norm aggregation
\begin{equation}\label{eqn:PsiPnorm}
\Psi \approx \left(\sum_e \left|\psi_e\right|^p \right)^{\frac{1}{p}}
\end{equation}
with
\begin{equation} \label{eqn:psiDefi}
\psi_e = \frac{A}{A_e}\frac{\mathrm{d} \Phi}{\mathrm{d} \rho_e}
\end{equation}
Here, the factor $\frac{A}{A_e}$ was introduced to eliminate the dependence of the sensitivity on the element size, thereby providing a mesh-independent measure of hot spot intensity, termed area (volume)-normed sensitivity. Equation~\ref{eqn:PsiPnorm} approaches the maximum operator as $p\rightarrow\infty$. In the original work, $p=32$ is employed as a compromise between a sharp approximation of the maximum and numerical smoothness.

\subsection{First-order worst-case approximation} \label{subsec:worstcases}
The SHoSP formulation was originally motivated by suppressing localized peaks in the objective sensitivity field. Viewed through a first-order robustness lens, this penalization can be understood more specifically as controlling the worst possible deterioration of the nominal objective under a budgeted material-mass uncertainty. Such an uncertainty model represents local over- or under-realization of material within the design domain, as may arise from manufacturing inaccuracies or other material-distribution deviations. In the following, we show that this interpretation leads directly to the SHoSP-augmented objective through a first-order worst-case approximation.

Consider a small perturbation $\delta\rho_e$ in the physical density, which corresponds to a perturbation of the material mass associated with this element,
\begin{equation}
\delta M_e = A_e \delta\rho_e ,
\end{equation}
Thus, $\delta M_e$ represents a local material-mass perturbation. We assume that the admissible perturbations are bounded by a prescribed $\ell_r$-norm budget on the vector of material-mass perturbations,
\begin{equation}\label{eqn:materialContentUncertainty}
\mathcal U_r(\kappa) = \left\{\delta M:\|\delta M\|_r \le \kappa A \right\}
\end{equation}
Here $\kappa$ defines the size of the admissible material perturbation as a fraction of the design domain area (or volume), 
making it dimensionless and avoiding an explicit dependence on the element size or mesh resolution.
The exponent $r$ controls how the admissible perturbation budget is allowed to distribute spatially across elements. 
Smaller values of $r$, approaching $r=1$, allow the budget to concentrate more strongly on a limited number of elements, whereas increasing $r$ progressively permits more distributed perturbations. In the limiting case $r=\infty$, the constraint reduces to a bound on the maximum element-wise perturbation. In particular, for $r=1$, Eq.~\ref{eqn:materialContentUncertainty} becomes
\begin{equation} \label{eqn:deltaMe}
\sum_e |\delta M_e|=\sum_e A_e|\delta\rho_e| \leq \kappa A ,
\end{equation}
which corresponds to a $\ell_1$-type cumulative material-mass budget and allows the perturbation to concentrate strongly on a limited number of elements. The implications of other choices of $r$ are discussed further in Sec.~\ref{subsec:uncertaintyNorm}.

The nominal objective is regarded as a reduced function of the physical densities, $\Phi=\Phi(\bm\rho)$, its first-order Taylor expansion gives
\begin{equation}\label{eqn:firstOrderTaylor}
\Phi(\bm\rho+\delta\bm\rho)
= \Phi(\bm\rho) + \sum_e \frac{\mathrm d \Phi}{\mathrm d \rho_e} \delta\rho_e + \mathcal O(\|\delta\bm\rho\|^2).
\end{equation}
Using $\delta\rho_e=\delta M_e/A_e$, the first-order perturbation term can be written in terms of the material-mass perturbation as
\begin{equation} \label{eqn:firstOrderMaterialContent}
\sum_e \frac{\mathrm d \Phi}{\mathrm d \rho_e} \delta\rho_e = \sum_e \frac{1}{A_e}\frac{\mathrm d \Phi}{\mathrm d \rho_e}\delta M_e
\end{equation}
The worst first-order increase over the uncertainty set in Eq.~\ref{eqn:materialContentUncertainty} is therefore
\begin{equation}\label{eqn:worstCaseMaterialContent}
\max_{\|\delta\bm M\|_r\le \kappa A} \sum_e \frac{1}{A_e} \frac{\mathrm d \Phi}{\mathrm d \rho_e}\delta M_e
\end{equation}
According to Hölder's inequality, i.e., for two vectors \(\bm a\) and \(\bm b\) of the same dimension, it fulfills
\begin{equation}\label{eqn:holderInequality}
\bm a^T\bm b \le \|\bm a\|_p\|\bm b\|_r, \qquad \frac{1}{p}+\frac{1}{r}=1
\end{equation}
Thereby, using $a_e =\frac{1}{A_e}\frac{\mathrm d\Phi}{\mathrm d\rho_e}$ and $b_e=\delta M_e$, together with Eq.~\ref{eqn:materialContentUncertainty}, the maximum in Eq.~\ref{eqn:worstCaseMaterialContent} is bounded by
\begin{equation}
\label{eqn:holderMaterialContent}
\max_{\|\delta\bm M\|_r\le \kappa A} \sum_e \frac{1}{A_e}\frac{\mathrm d \Phi}{\mathrm d \rho_e} \delta M_e
\leq \left(\sum_e\left|\frac{1}{A_e}\frac{\mathrm d \Phi}{\mathrm d \rho_e}\right|^p\right)^{1/p}\left(\sum_e\left|\delta M_e\right|^r\right)^{1/r}
\leq \kappa A \left(\sum_e\left|\frac{1}{A_e}\frac{\mathrm d \Phi}{\mathrm d \rho_e}\right|^p\right)^{1/p}
\end{equation}
Equivalently,
\begin{equation}\label{eqn:holderPsi}
\kappa A\left(\sum_e\left|\frac{1}{A_e}\frac{\mathrm d \Phi}{\mathrm d \rho_e}\right|^p\right)^{1/p} 
= \kappa\left(\sum_e\left|\frac{A}{A_e}\frac{\mathrm d\Phi}{\mathrm d\rho_e}\right|^p\right)^{1/p} 
= \kappa\left(\sum_e\left|\psi_e\right|^p\right)^{1/p}
= \kappa \|\bm\psi\|_p
\end{equation}
Since the H\"older's bound is attainable, as demonstrated below by the corresponding worst-case perturbation, combining the first-order expansion in Eq.~\ref{eqn:firstOrderTaylor} with the worst-case perturbation term in Eq.~\ref{eqn:holderPsi} gives the first-order approximation of the worst-case objective,
\begin{equation}\label{eqn:firstOrderRobustObj}
\Phi^{(1)}=\Phi+\kappa \|\bm\psi\|_p
\end{equation}

This result shows that $\kappa$, introduced in Eq.~\ref{eqn:materialContentUncertainty} as the normalized size of the admissible perturbation set, reappears unchanged as the weight on the sensitivity aggregation in the augmented objective (Eq.~\ref{eqn:augObj}).
Thus, as hypothesized but not proven in~\cite{sigmund2026}, $\kappa$ can be interpreted as the prescribed material-mass perturbation budget relative to the area or volume of the design domain. In the worst-case setting, this budget is distributed adversarially, i.e., in the admissible manner that produces the largest first-order deterioration of the nominal objective.

The appearance of the normalized sensitivity $\psi_e$ follows directly from the use of material-mass perturbations. Since $M_e=A_e\rho_e$,
\begin{equation}
\frac{\mathrm d\Phi}{\mathrm d M_e}
=\frac{1}{A_e}\frac{\mathrm d\Phi}{\mathrm d\rho_e} \qquad \Rightarrow
\qquad \psi_e = A \frac{\mathrm d\Phi}{\mathrm d M_e} = \frac{A}{A_e}\frac{\mathrm d\Phi}{\mathrm d\rho_e}
\end{equation}
Thus, the SHoSP sensitivity is the objective sensitivity with respect to element-wise material mass, scaled by the total area or volume of the design domain.

\paragraph{\textbf{Worst-case perturbation.}}
The derivation above yields the maximum first-order deterioration under the prescribed material-mass perturbation budget. The perturbation that attains this maximum can also be obtained from the equality condition of H\"older's inequality. For $1<p<\infty$, equality in Eq.~\ref{eqn:holderInequality} requires
\begin{equation} \label{eqn:holderEqualityCond}
    |b_e|^r = C |a_e|^p, \qquad \forall e
\end{equation}
for some constant $C>0$, with $a_e$ and $b_e$ having the same sign. Hence,
\begin{equation} \label{eqn:holderEqualityCondVar1}
    b_e = C\,\mathrm{sign}(a_e)|a_e|^{\frac{p}{r}} = C\,\mathrm{sign}(a_e)|a_e|^{p-1}
\end{equation}
where $C>0$ and $p/r=p-1$ follows from $1/p+1/r=1$. Substituting
$a_e =\frac{1}{A_e}\frac{\mathrm d\Phi}{\mathrm d\rho_e}$ and $b_e=\delta M_e$ gives
\begin{equation} \label{eqn:holderEqualityCondPhy}
    \delta M_e^{\mathrm{wc}} = C\,\mathrm{sign}(\psi_e)|\psi_e|^{p-1},
\end{equation}
where the constant $C$ absorbs the common factor associated with $A$ and is determined by the perturbation budget. Enforcing
$\|\delta\bm M^{\mathrm{wc}}\|_r=\kappa A$ and using $(p-1)r=p$ gives
\begin{equation} \label{eqn:holderEqualityCondPhyVar1}
    C = \frac{\kappa A}{\left(\sum_i|\psi_i|^p\right)^{1/r}}
\end{equation}
Accordingly,
\begin{equation} \label{eqn:wcTheoretical}
    \delta\rho_e^{\mathrm{wc}} =
    \frac{\delta M_e^{\mathrm{wc}}}{A_e}
    = \kappa\frac{A}{A_e}\frac{\mathrm{sign}(\psi_e)|\psi_e|^{p-1}}
    {\left(\sum_i|\psi_i|^p\right)^{1/r}}
\end{equation}
This perturbation conceptually attains the maximum first-order deterioration $\kappa\|\bm\psi\|_p=\kappa\Psi$ derived above. However, since it is obtained solely from the global material-mass norm constraint, it does not necessarily satisfy the point-wise admissibility bounds on the physical densities. Although SHoSP tends to make the corresponding H\"older worst-case perturbation less spatially concentrated as sensitivity hot spots are suppressed, this does not guarantee point-wise admissibility. Thus, a bounded approximation of this formal worst-case perturbation is still needed for the finite-perturbation robustness assessment, which will be introduced in Sec.~\ref{subsec:robustnessTest}.

\subsection{Uncertainty-set geometry and alternative perturbation modes} \label{subsec:uncertaintyNorm}
The derivation above defines a family of first-order worst-case models through the dual exponents $p$ and $r$. Different choices of these exponents prescribe how the available material-mass perturbation may be distributed over the design domain.

The limiting case $r=1$ ($p=\infty$) corresponds to a fixed accumulated material-mass budget. In this case, positive and negative material-mass deviations do not cancel, and the worst perturbation can concentrate the available budget on the most sensitive elements. The dual penalty is therefore a maximum-type sensitivity measure. This setting is consistent with the original motivation of SHoSP~\cite{sigmund2026}, namely to suppress localized sensitivity hot spots. The value $p=32$ used in the original formulation is a smooth approximation of this maximum-type penalty and corresponds to $r=32/31$, which is close to the $r=1$ uncertainty mode.

Other values of $r$ correspond to different perturbation assumptions. For example, $r=2$ leads to an $\ell_2$-type sensitivity aggregation and represents a more distributed material-mass perturbation. In contrast, $r=\infty$ bounds the element-wise perturbation magnitude and leads to an $\ell_1$-type aggregation of the sensitivity magnitudes. These alternatives may be relevant when the dominant uncertainty is expected to be spatially distributed or uniformly bounded rather than concentrated near a few critical elements.

In the present work, we retain the original SHoSP setting with $p=32$. This near-$\ell_1$ material-mass uncertainty is particularly aligned with the goal of identifying and suppressing localized sensitivity hot spots, while preserving a differentiable approximation of the maximum sensitivity. Other uncertainty modes can be interpreted within the same first-order framework and may be explored for applications in which a different perturbation structure is more appropriate.

\paragraph{\textbf{Perturbation space.}}
As established above, the uncertainty model adopted in the present formulation is defined directly in the physical-density space, or equivalently in terms of the element-wise material-mass perturbations $\delta\bm M$. Conceptually, the perturbation could also be prescribed in the underlying design-variable space $\bm x$, as adopted, for example, in the FOSM formulation of Kriegesmann and Lüdeker~\cite{kriegesmann2019robust}. In that case, a design-variable perturbation $\delta\bm x$ is mapped to the physical density through the filtering and projection operations, and, to the first order
\begin{equation}
    \delta\bm\rho
    \approx
    \frac{\partial\bm\rho}{\partial\bm x}\,\delta\bm x
\end{equation}
Consequently, the magnitude and spatial distribution of the resulting physical perturbation depend not only on the prescribed perturbation in $\bm x$, but also on the design parametrization and the associated filtering and projection operators. In particular, a localized perturbation in the design-variable space does not generally remain localized in the physical-density field, and its physical effect may vary with the filter radius, projection parameters, and the current design. By defining the uncertainty directly in $\bm\rho$, the present formulation instead places the perturbation budget on the material representation directly used in the constitutive interpolation and structural analysis. The resulting parameter $\kappa$ therefore retains a direct interpretation as a normalized material-mass perturbation budget, without its meaning being mediated by the design-to-physical mapping. This choice is also more closely aligned with the manufacturing interpretation considered here, where the uncertainty represents deviations in the realized material distribution rather than perturbations of the optimization parametrization.

\subsection{Adjoint sensitivity analysis for SHoSP} \label{subsec:adjoint}
The adjoint sensitivity analysis required for optimizing the augmented objective was derived in the original SHoSP study~\cite{sigmund2026}. Here, we state the final forms used in the present implementation and make explicit the auxiliary adjoint systems introduced by the penalization term. For completeness, a compact derivation is provided in Appendix~\ref{apdx:Adjoint}.

The derivative of the augmented objective is written as
\begin{equation}\label{eqn:augObjSensDef}
\frac{\mathrm{d}\tilde{\Phi}}{\mathrm{d}\rho_e} = \frac{\mathrm{d}\Phi}{\mathrm{d}\rho_e} + \kappa \frac{\mathrm{d}\Psi}{\mathrm{d}\rho_e}.
\end{equation}
Since the hot spot measure is constructed from the sensitivity, its derivative can be expressed as
\begin{equation}\label{eqn:penaSens}
\frac{\mathrm{d}\Psi}{\mathrm{d}\rho_e} = \sum_i \frac{\partial \Psi}{\partial \psi_i} \frac{\mathrm{d} \psi_i}{\mathrm{d} \rho_e} 
\end{equation}
To facilitate the notation, we define the aggregation weight
\begin{equation}\label{eqn:wDefi}
w_i := \frac{\partial \Psi}{\partial \psi_i} = \Psi^{1-p} \left|\psi_i\right|^{p-1} \mathrm{sign}\left(\psi_i\right)
\end{equation}

\paragraph{\textbf{Compliant mechanism design}.}
For compliant mechanism design, the sensitivity of the augmented objective with respect to the physical density is obtained as
\begin{equation}\label{eqn:dAugObjSensCoMe}
    \frac{\mathrm{d}\tilde{\Phi}} {\mathrm{d}\rho_e} = 
    \bm{\lambda}^{T} \frac{\partial \bm{K}}{\partial \rho_e} \bm{U} + 
    \kappa \left(\frac{w_e A}{A_e}\bm{\lambda}^{T} \frac{\partial^2 \bm{K}}{\partial\rho^2_e} \bm{U} + 
    \bm{\xi}^T\frac{\partial \bm{K}}{\partial \rho_e}\bm{U} + \bm{\eta}^T\frac{\partial \bm{K}}{\partial \rho_e}\bm{\lambda} \right)
\end{equation}
Here, two additional adjoint vectors, $\bm{\xi}$ and $\bm{\eta}$, are introduced. They are obtained by solving
\begin{equation}\label{eqn:dAugObjSensCoMeAddiSys1}
    \bm{K}\bm{\xi} = \bm{b}_{\lambda}, \qquad 
    \bm{b}_{\lambda}=-\sum_i \frac{w_iA}{A_i}\frac{\partial \bm{K}}{\partial \rho_i}\bm{\lambda}
\end{equation}
and
\begin{equation}\label{eqn:dAugObjSensCoMeAddiSys2}
    \bm{K}\bm{\eta} = \bm{b}_{U}, \qquad 
    \bm{b}_U=-\sum_i\frac{w_iA}{A_i}\frac{\partial \bm{K}}{\partial \rho_i}\bm{U}
\end{equation}
The same stiffness matrix as in the state and nominal adjoint equations is used in both auxiliary systems. Hence, SHoSP introduces additional right-hand sides for the same stiffness matrix, but no additional system matrices.

\paragraph{\textbf{Compliance minimization}.}
For compliance minimization, the problem is self-adjoint. With the sign convention adopted in Sec.~\ref{sec:TO}, the nominal adjoint satisfies $\bm{\lambda}=-\bm{U}$. Substituting this relation into the compliant-mechanism expressions makes the two auxiliary adjoint systems linearly dependent, so that only one additional adjoint solve is required. 
For convenience, we redefine the remaining auxiliary adjoint with the opposite sign, such that
\begin{equation}\label{eqn:dAugObjSensCoMiAddiSys}
    \bm{K}\bm{\eta} = \bm{b}, \qquad 
    \bm{b}=\sum_e \frac{w_eA}{A_e}\frac{\partial \bm{K}}{\partial \rho_e}\bm{U},
\end{equation}
the sensitivity of the augmented compliance objective becomes
\begin{equation}\label{eqn:dAugObjSensCoMi}
    \frac{\mathrm{d}\tilde{\Phi}} {\mathrm{d}\rho_e} = 
    -\bm{U}^{T} \frac{\partial \bm{K}}{\partial \rho_e} \bm{U} + 
    \kappa \left(2\bm{\eta}^T \frac{\partial \bm{K}}{\partial \rho_e}\bm{U} - \frac{w_eA}{A_e}\bm{U}^{T} \frac{\partial^2 \bm{K}}{\partial\rho^2_e} \bm{U} \right)
\end{equation}
All sensitivities above are defined with respect to the physical densities. Before being supplied to the optimizer, they are propagated through the projection and filtering operations using the chain rule, as described in Sec.~\ref{sec:TO}.

\section{Mechanical interpretation of sensitivity hot spots} \label{sec:stressSensHotSpots}

A notable observation in the original SHoSP study~\cite{sigmund2026} is that penalizing sensitivity hot spots often alleviates stress concentrations, even though no stress measure is included in the augmented objective. Similar behavior is also observed in compliant mechanism design, where the localized hinge-like deformation patterns are replaced by more distributed compliant regions, see Fig.~\ref{fig:Recap_SHoSP}. This section examines the mechanical origin of these effects. For compliance minimization, we show that objective sensitivity hot spots are closely related to local stress-energy concentrations. For compliant mechanism design, the corresponding hot spots are governed by the interaction between the state and adjoint stress fields, and therefore indicate localized force-transmission mechanisms rather than a single stress measure.

\subsection{Compliance minimization problems}
From Eq.~\ref{eqn:ComplianceSensitivity}, and according to the SIMP material assumption, the element-wise objective sensitivity with respect to the physical density in compliance minimization can be written as
\begin{equation} \label{eqn:psi_compliance_energy}
    \frac{\mathrm d \Phi}{\mathrm d \rho_e}
    = -\bm{U}^{T} \frac{\partial \bm{K}}{\partial \rho_e} \bm{U} 
    = -\frac{\mathrm d E_e}{\mathrm d \rho_e}\bm{U}_e^{T}\bm{K}_e^{unit}\bm{U}_e
\end{equation}
where $\bm{U}_e$ is the element displacement vector and $\bm{K}_e^{unit}$ is the element stiffness matrix corresponding to unit Young's modulus. With the SIMP interpolation in Eq.~\ref{eqn:SIMP}, and neglecting the small stiffness $E_{\min}$, one has
\begin{equation} \label{eqn:dEdrho}
    \frac{\mathrm d E_e}{\mathrm d \rho_e} = q (E_0-E_{\min})\rho_e^{q-1} \approx q E_0\rho_e^{q-1}
\end{equation}
The unit-modulus element strain energy can be expressed as
\begin{equation}\label{eqn:unit_energy}
    \bm{U}_e^{T} \bm{K}_e^{unit} \bm{U}_e = \frac{1}{E_0}
    \int_{\Omega_e} \bm{\varepsilon}_e^{T} \bm{D}_0 \bm{\varepsilon}_e \,\mathrm d\Omega
\end{equation}
where $\bm{\varepsilon}_e$ denotes the element strain field and $\bm{D}_0$ is the constitutive matrix corresponding to $E_0$.

To relate this expression to stress, we write the evaluated stress field in the form
\begin{equation} \label{eqn:stress_cmpt}
    \bm{\sigma}_e = f_{\sigma}(\rho_e) \bm{D}_0 \bm{\varepsilon}_e
\end{equation}
Here, $f_{\sigma}$ denotes the stress interpolation function. This interpolation need not coincide with the SIMP stiffness interpolation, since the latter is introduced for stiffness penalization and may excessively suppress stresses in intermediate-density regions. Using Eq.~\ref{eqn:stress_cmpt}, the unit-modulus strain energy can be equivalently written as
\begin{equation}
\label{eqn:unit_energy_stress}
    \bm{U}_e^{T} \bm{K}_e^{unit} \bm{U}_e = \int_{\Omega_e} \frac{1}{f_{\sigma}^2 E_0} \bm{\sigma}^{T} \bm{D}_0^{-1} \bm{\sigma} \,\mathrm d\Omega
\end{equation}
Substituting Eqs.~\ref{eqn:dEdrho} and \ref{eqn:unit_energy_stress} into Eq.~\ref{eqn:psi_compliance_energy} gives
\begin{equation}\label{eqn:psi_stress_compliance}
    \frac{\mathrm d \Phi}{\mathrm d \rho_e} \approx -q
    \frac{\rho_e^{q-1}}{f_{\sigma}^2(\rho_e)} \int_{\Omega_e} \bm{\sigma}^{T} \bm{D}_0^{-1} \bm{\sigma} \,\mathrm d\Omega
\end{equation}
Equation~\ref{eqn:psi_stress_compliance} relates compliance sensitivity hot spots to a density-weighted local stress-energy measure, with the weighting determined by the adopted stress interpolation.

For the stress evaluation used in this work, we adopt the $\epsilon$-relaxation interpolation
\begin{equation}\label{eqn:epsilonRelax}
    f_{\sigma}(\rho_e) = \frac{\rho_e}{\epsilon(1-\rho_e)+\rho_e},
\end{equation}
which is widely used in stress-constrained topology optimization to obtain a stable stress representation in intermediate-density regions~\cite{cheng1997varepsilon,da2019stress,da2021three}. Following these studies, we use $\epsilon=0.2$. Substituting Eq.~\ref{eqn:epsilonRelax} into Eq.~\ref{eqn:psi_stress_compliance} and considering $q=3$ yields
\begin{equation}\label{eqn:psi_stress_compliance_simp}
    \frac{\mathrm d \Phi}{\mathrm d \rho_e} 
    \approx - q\underbrace{\left(\epsilon(1-\rho_e)+\rho_e\right)^2}_{c_{\rho}}
    \int_{\Omega_e} \bm{\sigma}^{T} \bm{D}_0^{-1} \bm{\sigma}\,\mathrm d\Omega .
\end{equation}

\begin{figure}
    \centering
    \includegraphics[width=1.0\linewidth]{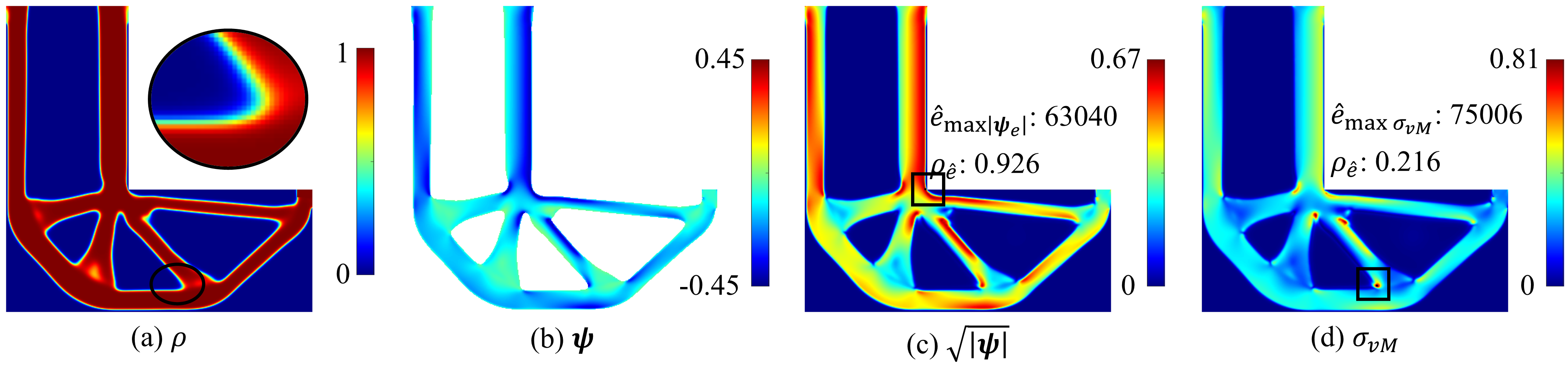}
    \caption{Relation between sensitivity-derived fields and von Mises stress in a compliance minimization example. The field $\sqrt{|\psi|}$ is included only as a stress-amplitude-like visualization of the sensitivity-derived measure. The comparison shows that sensitivity hot spots and von Mises stress concentrations are closely related but not identical, due to density weighting and different scalarizations of the local stress state.}
    \label{fig:sensitivity_stress_relation}
\end{figure}

Equation~\ref{eqn:psi_stress_compliance_simp} shows that a compliance sensitivity hot spot corresponds to a large density-weighted local stress-energy measure. In nearly solid regions, $c_\rho \approx 1$, suppressing sensitivity peaks tends to reduce local stress concentrations in the energy norm. To relate this result to the von Mises stress used throughout the paper, we note that, for isotropic linear elasticity, the stress-energy measure contains both deviatoric and volumetric contributions, whereas the von Mises stress depends only on the deviatoric component. Their spatial localization may therefore be similar for particular stress states, but no general one-to-one correspondence is implied. The detailed relation is summarized in Appendix~\ref{apdx:stressMeasures}. The density-dependent factor $c_\rho$ further modifies this relation in intermediate-density regions. Consequently, even the stress-amplitude-like field $\sqrt{|\psi|}$ does not generally coincide with the von Mises stress distribution, as illustrated in Fig.~\ref{fig:sensitivity_stress_relation}. SHoSP should therefore be regarded as indirectly suppressing localized stress-energy concentrations rather than as an equivalent form of stress-oriented topology optimization.

\subsection{Compliant mechanism design problems}
For compliant mechanism design, a similar derivation can be performed to examine the mechanical interpretation of the objective sensitivity. However, unlike compliance minimization, the resulting sensitivity cannot be associated with a single stress field, since the objective sensitivity is governed by both the state and adjoint responses.

Starting with
\begin{equation} \label{eqn:mechanism_sens_unit}
    \frac{\mathrm d \Phi}{\mathrm d \rho_e} 
    = \frac{\mathrm d E_e}{\mathrm d \rho_e} \bm{\lambda}_e^{T} \bm{K}_e^{unit} \bm{U}_e
\end{equation}
and introducing the state and adjoint stress fields,
\begin{equation}\label{eqn:state_adjoint_stress}
    \bm{\sigma}_U = f_{\sigma}(\rho_e)\bm{D}_0
    \bm{\varepsilon}_U,
    \qquad
    \bm{\sigma}_{\lambda} = f_{\sigma}(\rho_e)\bm{D}_0\bm{\varepsilon}_{\lambda}
\end{equation}
where $\bm{\varepsilon}_{U}$ and $\bm{\varepsilon}_{\lambda}$ denote the strain fields associated with the state displacement field ($\bm U$) and the adjoint displacement field ($\bm\lambda$), respectively. Equation~\ref{eqn:mechanism_sens_unit} can be rewritten in a stress-based form. Following the same procedure adopted for compliance minimization and employing the $\varepsilon$-relaxation stress interpolation, one obtains
\begin{equation}\label{eqn:psi_stress_mechanism}
    \frac{\mathrm d \Phi}{\mathrm d \rho_e} \approx
    q (\epsilon(1-\rho_e)+\rho_e)^2 \int_{\Omega_e} \bm{\sigma}_{\lambda}^{T} \bm{D}_0^{-1} \bm{\sigma}_{U}d\Omega
\end{equation}

\begin{figure}
    \centering
    \includegraphics[width=1.0\linewidth]{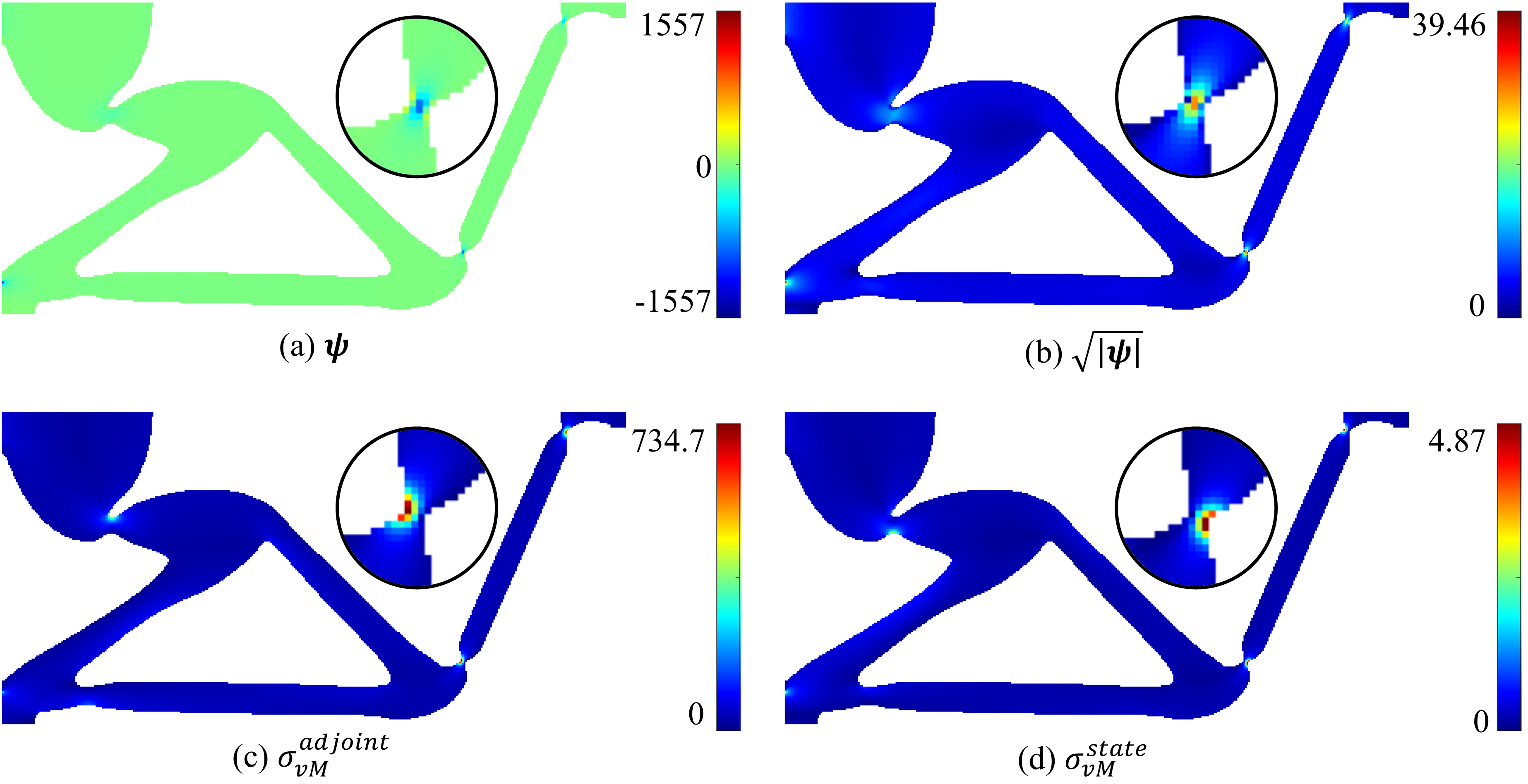}
    \caption{Sensitivity-derived fields and state/adjoint stress fields for the force inverter design by standard deterministic topology optimization. The figure compares the normalized objective sensitivity $\psi$, the visualization field $\sqrt{|\psi|}$, and the von Mises stresses associated with the adjoint field $\bm{\lambda}$ and the state field $\bm U$. 
    }
    \label{fig:problemExposition_CoMe}
\end{figure}

Equation~\ref{eqn:psi_stress_mechanism} shows that, in non-self-adjoint compliant mechanism design, sensitivity hot spots are not associated with a single stress field. Instead, they arise from the local interaction between the state and adjoint stress fields. This distinction is illustrated in Fig.~\ref{fig:problemExposition_CoMe}. The state and adjoint von Mises stress fields exhibit pronounced concentrations around the hinge region, but their peaks are located on different sides of the hinge. The sensitivity hot spot appears in the region where the two fields interact most strongly, rather than coinciding with either individual stress peak. Thus, unlike compliance minimization, the sensitivity distribution cannot be explained from the state stress or the adjoint stress alone.

This interaction-based interpretation explains why SHoSP removes localized hinge-like features in compliant mechanisms. In deterministic designs, the required force transmission and displacement inversion are often achieved through a small compliant hinge. Such a region carries a disproportionate part of the state-adjoint interaction and therefore appears as a sensitivity hot spot. Penalizing this hot spot does not directly minimize the state stress. Instead, it discourages the optimizer from concentrating the transmission mechanism in a small region. To maintain the required output performance, the deformation and force-transfer paths are redistributed over a larger compliant region, leading to the distributed hinge patterns observed in the original SHoSP study and in the examples below. The associated reduction in state-stress concentration is therefore a by-product of suppressing the localized transmission mechanism, rather than the result of an explicit stress objective.

The compliance and compliant-mechanism cases, therefore, give two different mechanical interpretations of sensitivity hot spots. In compliance minimization, they are closely related to density-weighted local stress-energy concentrations. In compliant mechanism design, they identify localized regions of strong state-adjoint interaction that dominate force transmission and output performance. In both cases, sensitivity hot spots indicate localized structural features on which the global response depends disproportionately. Suppressing them improves structural regularity and robustness, although the specific manifestation may be stress reduction, hinge redistribution, or a more distributed load-transfer mechanism.


\section{Implementation and scalability}\label{sec:implScale}
The preceding sections established the first-order worst-case approximation and mechanical interpretations of SHoSP. 
This section discusses practical aspects relevant to its implementation, extension, and scalable application.
We first summarize the computational overhead implied by the adjoint sensitivity analysis in Sec.~\ref{subsec:adjoint}. Then, we discuss the applicability of SHoSP to representative extended formulations, including cases where the objective aggregation or the admissible design space is modified. Finally, we comment on the practical choice of the penalization strength.

\subsection{Computational overhead} \label{subsec:cmpt}
Compared with standard deterministic topology optimization, the additional computational cost of SHoSP originates mainly from the auxiliary adjoint problems required by the hot-spot penalization term. As shown in Sec.~\ref{subsec:adjoint}, two additional linear system solves are required for compliant mechanism design, see Eqs.~\ref{eqn:dAugObjSensCoMeAddiSys1} and~\ref{eqn:dAugObjSensCoMeAddiSys2}, whereas one additional system is required for compliance minimization, see Eq.~\ref{eqn:dAugObjSensCoMiAddiSys}. These systems increase the number of RHSs, but they use the same stiffness matrix as the state equation and, for compliant mechanism design, the original adjoint equation. 

For 2D and small-scale 3D problems, direct solvers such as Cholesky factorization are often sufficient for the symmetric positive-definite systems considered here. In this case, the factorized stiffness matrix can be reused for all RHSs within the same optimization step. Since the optimization is performed on a fixed mesh, the sparsity pattern also remains unchanged, allowing the symbolic factorization to be reused while only the numerical factorization is updated. SHoSP naturally benefits from these standard reuse strategies.

For large-scale 3D problems, iterative solvers are usually required. The use of a common stiffness matrix is again advantageous, since the same matrix-specific preconditioner can be reused for the state, nominal adjoint, and auxiliary adjoint systems. In large-scale topology optimization on fixed Cartesian meshes, geometric multigrid preconditioned conjugate gradient solvers (GMGCG) combined with matrix-free sparse matrix-vector multiplication have been widely used because of their favorable convergence and memory efficiency~\cite{aage2015topology,wu2016system,traff2023simple,wang2025efficient}. In this work, we use an in-house matrix-free GMGCG implementation, denoted by \emph{mfGMGCG}, based on~\cite{wang2025efficient,wang2025sgldbench}, for the large-scale 3D examples.

Warm-starting the auxiliary adjoint systems deserves a separate comment. For the standard state equation, the external force vector is fixed, and the effectiveness of warm starts mainly relies on the gradual change of the stiffness matrix between consecutive optimization steps. For the SHoSP auxiliary systems, the situation is less straightforward because the right-hand sides are not fixed loads. Specifically, they are reassembled at each optimization step from the current state and adjoint fields, the sensitivity weights, and the density-dependent stiffness derivatives. One might therefore expect these systems to be less compatible with warm starts. In practice, however, the auxiliary right-hand sides are induced by the mechanical response and evolve coherently with the design. In the numerical examples considered here, the auxiliary systems remain well-suited for warm starts, with iteration counts generally no larger and often slightly smaller than those of the corresponding state and nominal adjoint solves.

\subsection{Applicability to extended formulations} \label{subsec:extension}
The SHoSP term is constructed from the sensitivity of the nominal objective. This makes the formulation directly applicable to extended density-based topology optimization problems, provided that the reduced objective sensitivity is available. However, the effect of the penalization is mediated by the formulation into which it is inserted. We consider two representative extensions of compliance minimization. The first is the multiple-load case, where separate load-specific SHoSP-augmented responses are treated within a min–max formulation~\cite{james2009structural}. The second is the local volume constrained case, often used for porous infill design~\cite{wu2017infill,dou2020projection}, where the admissible design space is restricted by additional material-distribution constraints. These two cases illustrate how SHoSP behaves when either multiple competing structural responses are considered or the feasible design space is further constrained.

\paragraph{\textbf{Multiple loading conditions.}}
For problems involving several prescribed loading conditions, we adopt a min--max formulation to avoid prescribing relative weights between the individual load cases and to control the worst load-specific SHoSP-augmented compliance. Applying SHoSP independently to each loading condition gives
\begin{equation}
\begin{aligned}
\min_{\bm{x},\,z} \quad & z, \\
\text{s.t.}\quad
& \bm{K}(\bm{\rho})\bm{U}_i=\bm{F}_i, && i=1,\ldots,M, \\
& \Phi_i(\bm{\rho})+\kappa\Psi_i(\bm{\rho})-z\le 0, && i=1,\ldots,M, \\
& g(\bm{\rho})\le 0, \\
& \bm{\rho} = \mathcal{P}\!\left(\mathcal{F}(\bm{x},r),\beta,\eta\right), \\
& 0\le x_e\le1, && \forall e .
\end{aligned}
\label{eq:multiLoads}
\end{equation}
where $M$ is the number of load cases, and $\bm U_i$ denotes the displacement vector of the $i$-th load case $\bm F_i$. $\Phi_i$ and $\Psi_i$ are given by Eqs.~\ref{eqn:Compliance} and~\ref{eqn:PsiPnorm}, respectively. The auxiliary variable $z$ bounds the SHoSP-augmented compliance of all load cases, while the hot spot measure is evaluated independently for each load-specific sensitivity field. Consequently, the governing load case, or multiple simultaneously active load cases, is determined directly by the optimization. This formulation also demonstrates that SHoSP-augmented responses can be incorporated into optimization constraints, rather than being restricted to a single augmented objective.

\paragraph{\textbf{Local volume constraints.}}
A different extension is obtained by keeping the nominal objective unchanged while restricting the admissible design space. A representative example is topology optimization with local volume constraints, which is often used to generate porous infill designs. Such constraints limit the amount of material allowed within a prescribed neighborhood of each element and can be written conceptually as
\begin{equation} \label{eqn:LVF}
    g_e(\bm\rho) = \bar{\rho}_e - V_{e0} \leq 0
\end{equation}
where $\bar{\rho}_e$ is the local averaged density and $V_{e0}$ is the prescribed local upper bound, where the locality is denoted by the effect radius $R_e$. In practice, the many local constraints are often aggregated into a differentiable global constraint (cf.~\cite{wu2017infill}). Since SHoSP is constructed from the nominal objective sensitivity, these additional constraints do not change the definition of the hot-spot measure or the associated adjoint sensitivity analysis.

The local volume constraint nevertheless affects how SHoSP can act on the design. Suppressing sensitivity hot spots tends to redistribute structural responsibility away from overly localized regions, whereas a local volume constraint restricts how much material can be placed or rearranged within each neighborhood and globally. The optimizer, therefore, has less freedom to relieve sensitivity localization than in an unconstrained solid-void formulation. As a result, the SHoSP effect may become more problem-dependent: local volume constraint reduces the available redistribution freedom, so the effect of a given $\kappa$ may differ from the standard formulation. This makes porous infill design a useful test case for the proposed formulation, since the first-order interpretation remains valid while the feasible material redistribution is strongly constrained.

\subsection{Practical choice of penalization strength} \label{subsec:penaltyStrength}

The SHoSP term modifies the augmented objective, thereby changing the descent direction provided to the optimizer. Its effect is consequently determined not only by the definition of the hot-spot measure, but also by when the penalization is activated and how strongly it is weighted relative to the nominal objective. This is important because topology optimization is path dependent: the main structural layout is usually established during the early iterations, whereas later iterations mainly refine boundaries, remove residual gray regions, and adjust local features. If SHoSP is introduced too weakly or too late, it may only regularize an already established topology. If it is introduced too strongly or too early, it may interfere with the formation of the intended structural response.

The appropriate activation strategy is problem dependent. For compliance minimization, the topology evolution is usually governed by the formation of load-carrying paths, and activating SHoSP from the beginning is generally acceptable. Early activation may even be beneficial when the aim is to influence the formation of less localizing load paths, similar to stress-oriented formulations where early-stage topology evolution is often decisive for stress reduction. For compliant mechanism design, however, the early iterations are often dominated by the search for a feasible force-transmission mechanism and, in inverter-type problems, by the correction of the output displacement direction. 
At this stage, the nominal output may be very small or even change sign, so large sensitivity peaks do not necessarily indicate an undesirable localized mechanism. Instead, they may be part of the transient formation of a viable transmission path.
Therefore, in the compliant mechanism examples considered in this work, the SHoSP term is activated only after the first update of the projection parameter $\beta$, allowing a preliminary mechanism to form before sensitivity localization is penalized. In practice, the activation is separated from the projection update itself, so that the optimizer is not subjected to simultaneous changes in both the projection sharpness and the augmented objective.

Although $\kappa$ has a clear interpretation as a dimensionless material-mass perturbation budget (cf. Sec.~\ref{subsec:worstcases}), the practical effect of a prescribed budget depends on the optimization formulation into which SHoSP is introduced. The same value of $\kappa$ has the same budget interpretation, but it may lead to different levels of nominal-performance trade-off depending on the scale of the objective, the magnitude and localization of $\Psi$, and the freedom available for material redistribution. In compliance minimization, the response is typically energy-like, and the design can often accommodate a moderate SHoSP penalty through smoother load paths. In porous infill formulations, local volume constraints restrict material redistribution, so an excessively large $\kappa$ may compete strongly with the admissible material-distribution constraints. In compliant mechanism design, a large $\kappa$ may over-regularize the compliant regions that are needed for the intended motion. Thus, the perturbation-budget interpretation provides a consistent meaning of $\kappa$, while the resulting optimization behavior and performance trade-off should be assessed in the context of the specific formulation.

A useful numerical diagnostic is the relative size of the SHoSP correction,
\begin{equation} \label{eqn:rkappa}
    r_{\kappa} = \frac{\kappa\Psi}{|\Phi|}.
\end{equation}
This quantity is introduced as a simple way to monitor whether the penalty is small, moderate, or dominant relative to the nominal objective. If $r_{\kappa}$ becomes too large, the augmented objective may be driven primarily by hot spot suppression rather than by the original design goal, which can lead to excessive nominal performance degradation or even optimization failure. Conversely, a very small $r_{\kappa}$ indicates that the SHoSP term is unlikely to have a visible effect. 
Monitoring $r_{\kappa}$ during continuation therefore provides a practical reference for interpreting the selected $\kappa$ and for comparing its effect across different optimization formulations.

\section{Results} \label{sec:results}
This section presents numerical examples organized around the interpretations and implementation aspects developed in the preceding sections. We first revisit two canonical examples from the original SHoSP study~\cite{sigmund2026} and perform a perturbation-based robustness assessment. Next, we compare SHoSP with stress-constrained topology optimization to clarify the relation and distinction between SHoSP and direct stress control. Then, we examine SHoSP in 3D and extended topology optimization formulations. Finally, three large-scale 3D examples are presented to demonstrate its scalability and computational efficiency.

\begin{figure}
    \centering
    \includegraphics[width=0.75\linewidth]{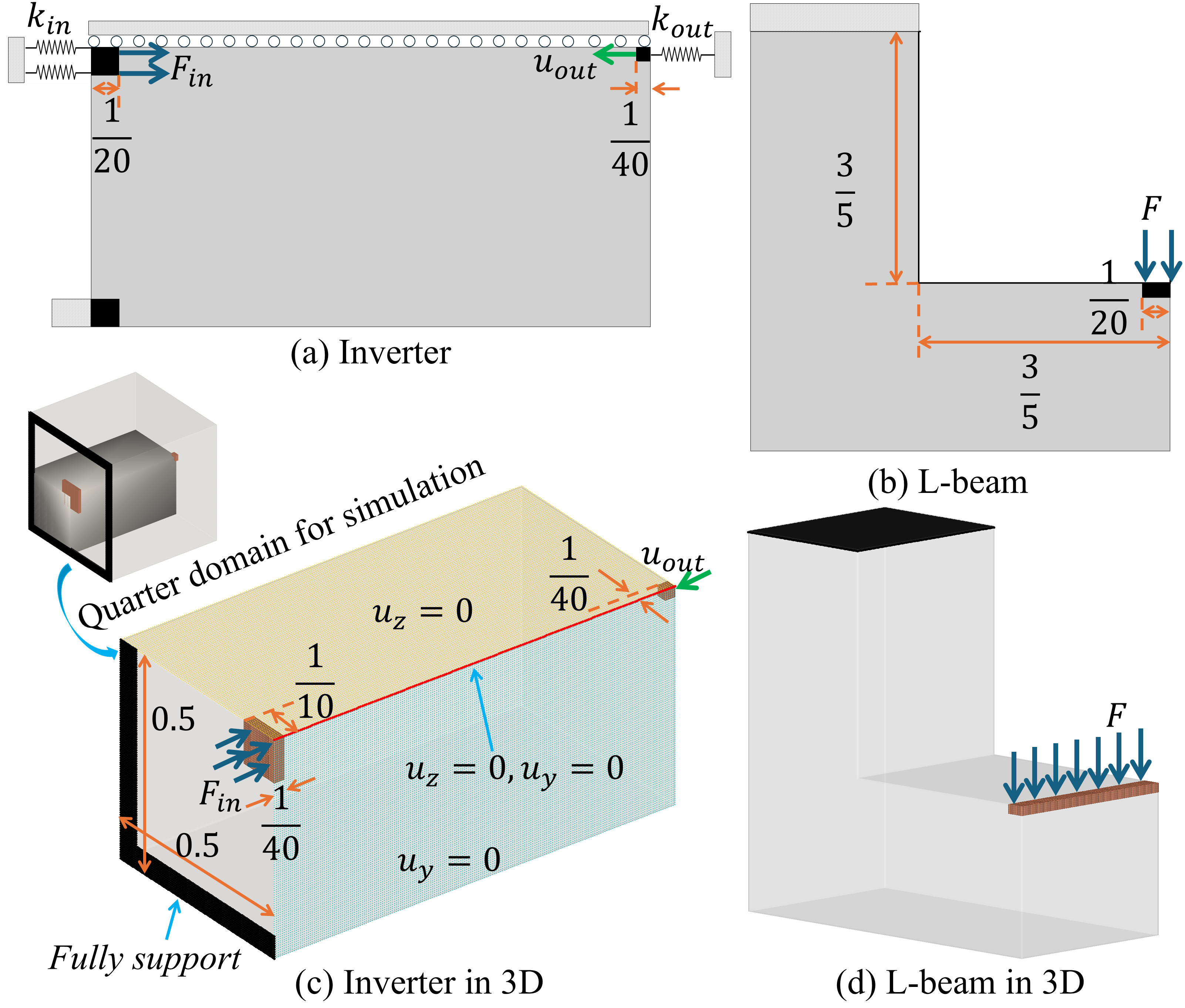}
    \caption{Problem descriptions of the 2D and 3D benchmark examples: the inverter for compliant mechanism design and the L-beam for compliance minimization. The input and output springs of the 3D inverter in (c) are omitted for clarity, but are applied in the same manner as in the 2D inverter in (a). The mesh resolutions are $400\times200$ for the 2D inverter, $400\times400$ for the 2D L-beam, $200\times100\times100$ for the 3D inverter, and $200\times100\times200$ for the 3D L-beam.}
    \label{fig:BenchmarkPDs}
\end{figure}

The method of moving asymptotes (MMA) is used as the optimizer. Unless otherwise stated, an outer move limit is set to 0.2 for compliant mechanism design and 0.1 for compliance minimization. Isotropic linear elasticity is assumed throughout, with Young's modulus $E=1.0$ and Poisson's ratio $\nu=0.3$. For compliant mechanism design, the input and output spring stiffnesses $k_{in}$ and $k_{out}$, defined in Fig.~\ref{fig:BenchmarkPDs}, are set to $1.0$ and $1.0\times10^{-3}$ for the 2D examples, and to $0.1$ and $1.0\times10^{-4}$ for the 3D examples, respectively. The values of $\kappa$ considered in the parametric studies are chosen to span from weak to pronounced SHoSP effects. The largest one is selected to retain a well-behaved optimization process, since excessively large values may cause the penalty term to dominate the nominal objective and impair convergence.

All numerical examples are optimized on fixed structured Cartesian meshes fitted to the computational design domains, with the maximum dimensions normalized to unity. For non-rectangular or non-box-shaped domains, only the active elements belonging to the prescribed design domain are included in the finite element analysis and in the volume fraction evaluation. For 3D examples, the matrix-free and multigrid operators are constructed directly on this active Cartesian discretization, and the resulting systems are solved using the in-house \emph{mfGMGCG} solver referred in Subsec.~\ref{subsec:cmpt}, with a relative convergence tolerance of $1.0\times10^{-5}$. All 3D computations are parallelized using OpenMP and performed on a desktop workstation equipped with an Intel Xeon W2235 CPU (6 cores/12 threads, 3.79 GHz) and four-channel DDR4 memory (64 GB).

The optimization process is terminated when either 400 optimization iterations are reached or the binary metric $s=\frac{1}{N_e}\sum_e 4\rho_e(1-\rho_e)$ falls below $1.0\times10^{-3}$. Unless otherwise specified, optimized designs are extracted using the density threshold $\rho\geq0.5$. Sensitivity and stress fields are evaluated element-wise on the structured meshes, with element stress defined as the stress state at the element centroid. Plane-stress conditions are assumed for the 2D examples. The canonical benchmark problems used throughout the following sections are summarized in Fig.~\ref{fig:BenchmarkPDs}.

\subsection{Verification and robustness assessment}\label{subsec:robustnessTest}
We begin with the 2D L-beam and inverter benchmarks used in the original SHoSP study~\cite{sigmund2026}, which correspond to compliance minimization and compliant mechanism design, respectively. These examples provide a controlled setting for revisiting the characteristic effects of SHoSP and for assessing the first-order worst-case interpretation developed in Sec.~\ref{subsec:worstcases}. For each benchmark, the influence of the penalization parameter $\kappa$ is demonstrated, and the stress and sensitivity trends observed on the structured meshes are further checked by independent body-fitted finite element analyses. The resulting designs are then used in the perturbation-based robustness assessment.

\paragraph{\textbf{Canonical 2D benchmarks.}}
The 2D L-beam and inverter benchmarks shown in Fig.~\ref{fig:BenchmarkPDs}(a,b) are considered here. The prescribed volume fractions are $V_0=0.3$ for the inverter and $V_0=0.4$ for the L-beam, both have a filter radius of $R=24h_e$. For the L-beam, the SHoSP parameter is set to $\kappa=0$, $0.05\%$, $0.1\%$, and $1\%$, where $\kappa=0$ corresponds to the standard deterministic formulation. For the inverter, $\kappa=0$, $0.03\%$, $0.05\%$, and $0.1\%$ are considered. Throughout the numerical examples, $\kappa$ is reported in percentage form according to its material-mass budget interpretation.

\begin{figure}
    \centering
    \includegraphics[width=1.0\linewidth]{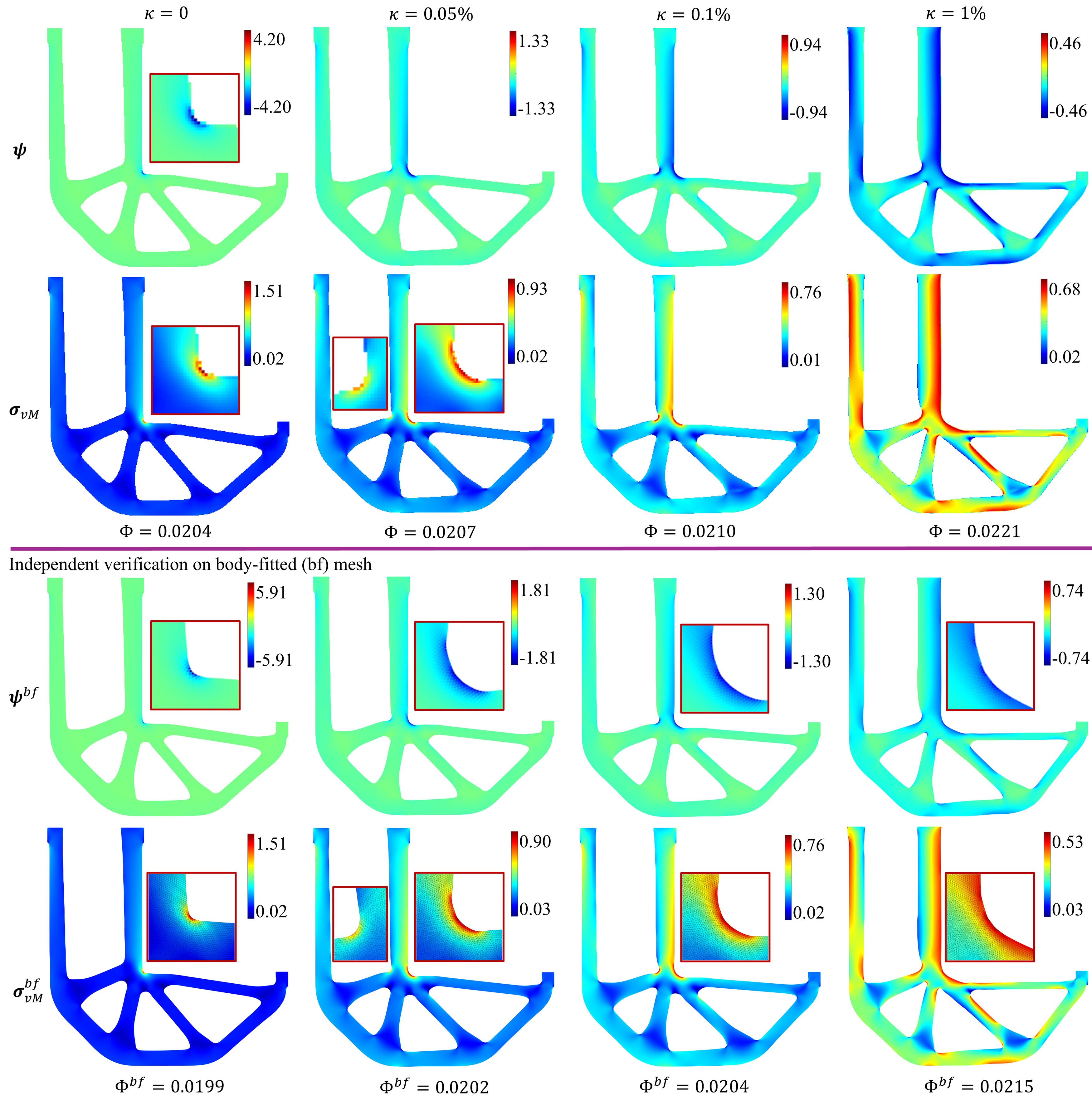}
    \caption{Effect of SHoSP on the 2D L-beam benchmark for different values of $\kappa$. The upper two rows show the normalized sensitivity field $\bm \psi$ and the von Mises stress field $\bm \sigma_{vM}$ evaluated on the Cartesian designs. The bottom row shows independent verification results on body-fitted meshes, including the body-fitted von Mises stress field $\bm \sigma_{vM}^{bf}$ and the reanalyzed objective value $\Phi^{bf}$.}
    \label{fig:2DLbeam}
\end{figure}

\begin{figure}
    \centering
    \includegraphics[width=1.0\linewidth]{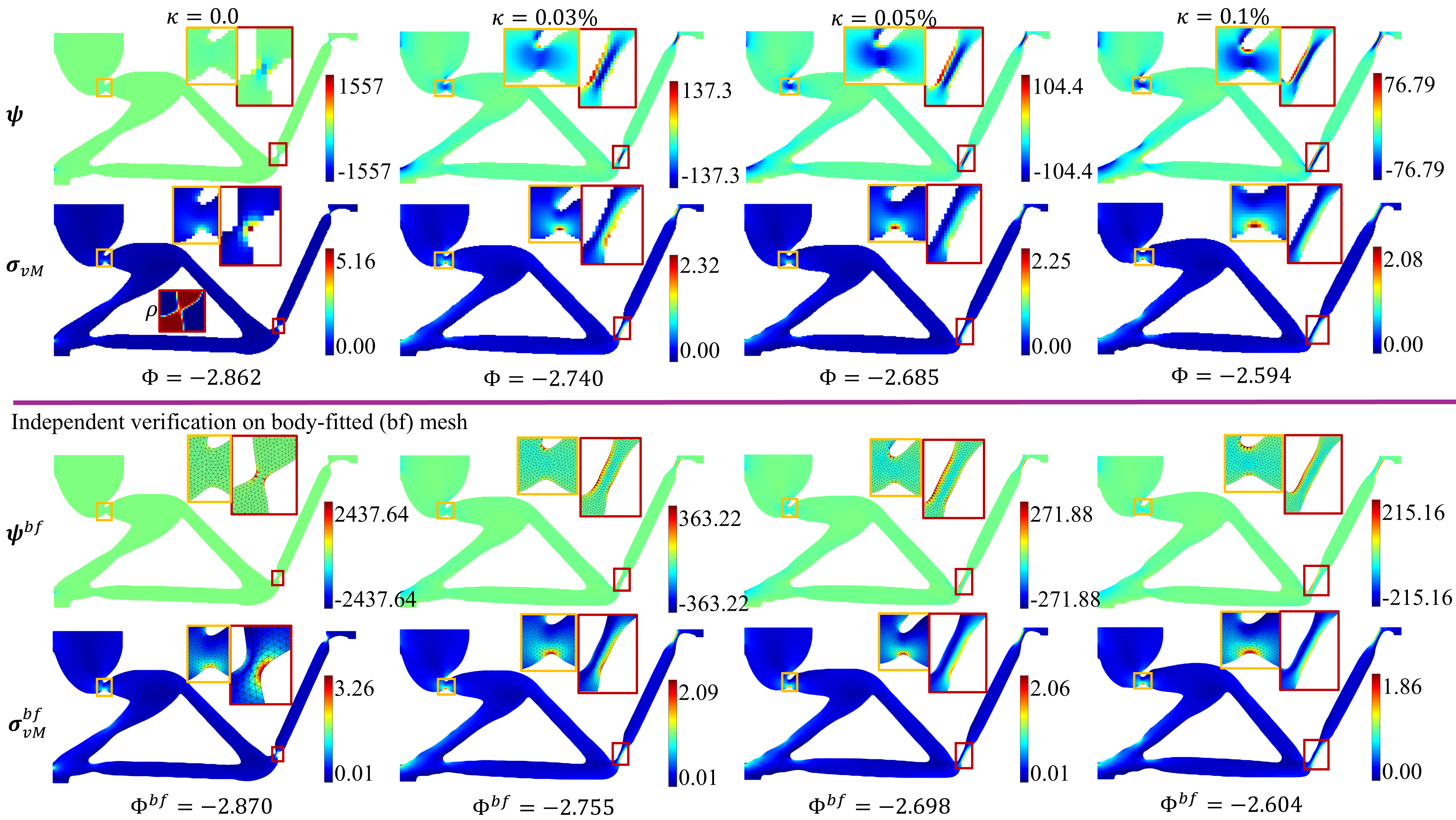}
    \caption{Effect of SHoSP on the 2D inverter benchmark for different values of $\kappa$. The results are organized using the same layout as in Fig.~\ref{fig:2DLbeam}, cf. its caption.}
    \label{fig:2Dinverter}
\end{figure}

The first two rows of Figs.~\ref{fig:2DLbeam} and~\ref{fig:2Dinverter} show the area-normalized sensitivity field ($\bm{\psi}$) and the von Mises stress field ($\bm{\sigma}_{vM}$) directly evaluated on the structured meshes. Increasing $\kappa$ progressively suppresses localized sensitivity hot spots in both benchmarks, while the associated stress or deformation localization is weakened. In the L-beam, the dominant hot spot appears near the re-entrant corner and is spatially correlated with the von Mises stress concentration, consistent with the relation between compliance sensitivities and density-weighted stress-energy concentrations discussed in Sec.~\ref{sec:stressSensHotSpots}. In the inverter, a point-wise correspondence with the state stress field is not expected, since the objective sensitivity is governed by the interaction between the state and adjoint responses. Nevertheless, suppressing the sensitivity hot spot weakens the localized hinge-like transmission mechanism, leading to a more distributed compliant region and a lower associated state-stress concentration.

To examine whether the observed sensitivity localization and stress redistribution persist beyond the fixed Cartesian discretization, an independent body-fitted finite element analysis is performed. Each optimized density field is first converted to nodal form, and the $\rho=0.5$ iso-contour is extracted and remeshed using linear triangular elements to obtain a body-fitted geometry. The corresponding boundary conditions are then transferred to the reconstructed finite element model. The objective, area-normalized elemental sensitivity field, and nodal von Mises stress field are subsequently re-evaluated using a standard finite element analysis pipeline\footnote{The open-source FEM package \emph{MiniFEM} at \href{https://github.com/PSLer/MiniFEM}{https://github.com/PSLer/MiniFEM} is used for this reanalysis.}. For the body-fitted sensitivity evaluation, the element area $A_e$ in the normalization is taken as the actual area of each triangular element, while the reference area $A$ is kept equal to the original design-domain area used in the structured-mesh analysis, thereby retaining the same normalization across the two discretizations.
The resulting elemental sensitivity and nodal stress fields are shown in the bottom rows of Figs.~\ref{fig:2DLbeam} and~\ref{fig:2Dinverter}.

The body-fitted reanalyses reproduce the nominal objective values with only small discrepancies: the relative difference $|\Phi^{bf}-\Phi|/|\Phi|$ remains below $3\%$ for the L-beam and below $1\%$ for the inverter. More importantly, the independently evaluated sensitivity and von Mises stress fields preserve the same tendency with increasing $\kappa$ as observed on the structured meshes. The peak von Mises stresses obtained on the body-fitted meshes are consistently lower than those evaluated on the structured meshes, indicating that the absolute peak stresses are affected by the density-based geometric representation and motivating their independent reassessment on the reconstructed solid geometries. For the L-beam, the larger stress discrepancy at $\kappa=1\%$ appears after the dominant stress concentration at the re-entrant corner has been substantially suppressed, leaving the reported peak stress more influenced by local differences between the optimized and reconstructed geometries. For the deterministic inverter, the larger discrepancy is associated with the narrow gray hinge that governs the original response, whose geometric representation is substantially altered by the $\rho=0.5$ contour extraction. Despite these differences in peak magnitude, all cases exhibit the same systematic reduction and redistribution of sensitivity and stress concentrations with increasing $\kappa$, confirming that the observed trends are associated with changes in the optimized structural layouts.

\paragraph{\textbf{Perturbation-based robustness assessment.}}
We next use the deterministic and SHoSP designs shown in Figs.~\ref{fig:2DLbeam} and~\ref{fig:2Dinverter} to assess the first-order worst-case interpretation proposed in Sec.~\ref{subsec:worstcases}. For a given optimized density field $\bm{\rho}$, admissible perturbations satisfy the material-mass budget and the density bounds, i.e., $\sum_e A_e |\delta\rho_e| \leq \kappa_{\mathrm{test}} A$ and $0 \leq \rho_e+\delta\rho_e \leq 1$. Here $\kappa_{\mathrm{test}}$ is the perturbation budget used in the assessment. The relative objective deterioration is measured by $\varphi(\delta\bm{\rho})=\frac{\Phi(\bm{\rho}+\delta\bm{\rho})-\Phi(\bm{\rho})}{|\Phi(\bm{\rho})|}$. The same definition is used for both compliance minimization and compliant mechanism design, so that a positive value of $\varphi$ indicates deterioration of the nominal objective. For each SHoSP design, $\kappa\Psi$ represents the maximum first-order objective degradation associated with the material-mass perturbation budget $\kappa$ considered during optimization, without imposing point-wise bounds on the perturbed densities.

To obtain a feasible finite perturbation under the density bounds, we employ a sensitivity-ranked adverse allocation consistent with the near-$\ell_1$ uncertainty mode associated with $p=32$.
Elements are sorted in descending order of $|\psi_e|$, and the perturbation is applied in the local adverse direction, i.e. the direction that increases the first-order variation of $\Phi$. The material-mass budget is then assigned successively to the sorted elements, subject to the density bounds, until the prescribed budget is exhausted. This procedure is used instead of the formal H\"older equality perturbation in Eq.~\ref{eqn:wcTheoretical}, since the latter may concentrate the budget on only a few highly sensitive elements and become ineffective once the bounds are enforced. For each SHoSP penalization level $\kappa$ adopted in Figs.~\ref{fig:2DLbeam} and~\ref{fig:2Dinverter}, the testing budget $\kappa_{\mathrm{test}}$ is varied from $0.25\kappa$ to $2.00\kappa$ in increments of $0.25\kappa$. The same testing budgets are applied to the corresponding deterministic design to provide a direct comparison under equal perturbation budgets.

\begin{figure}
    \centering
    \includegraphics[width=1.0\linewidth]{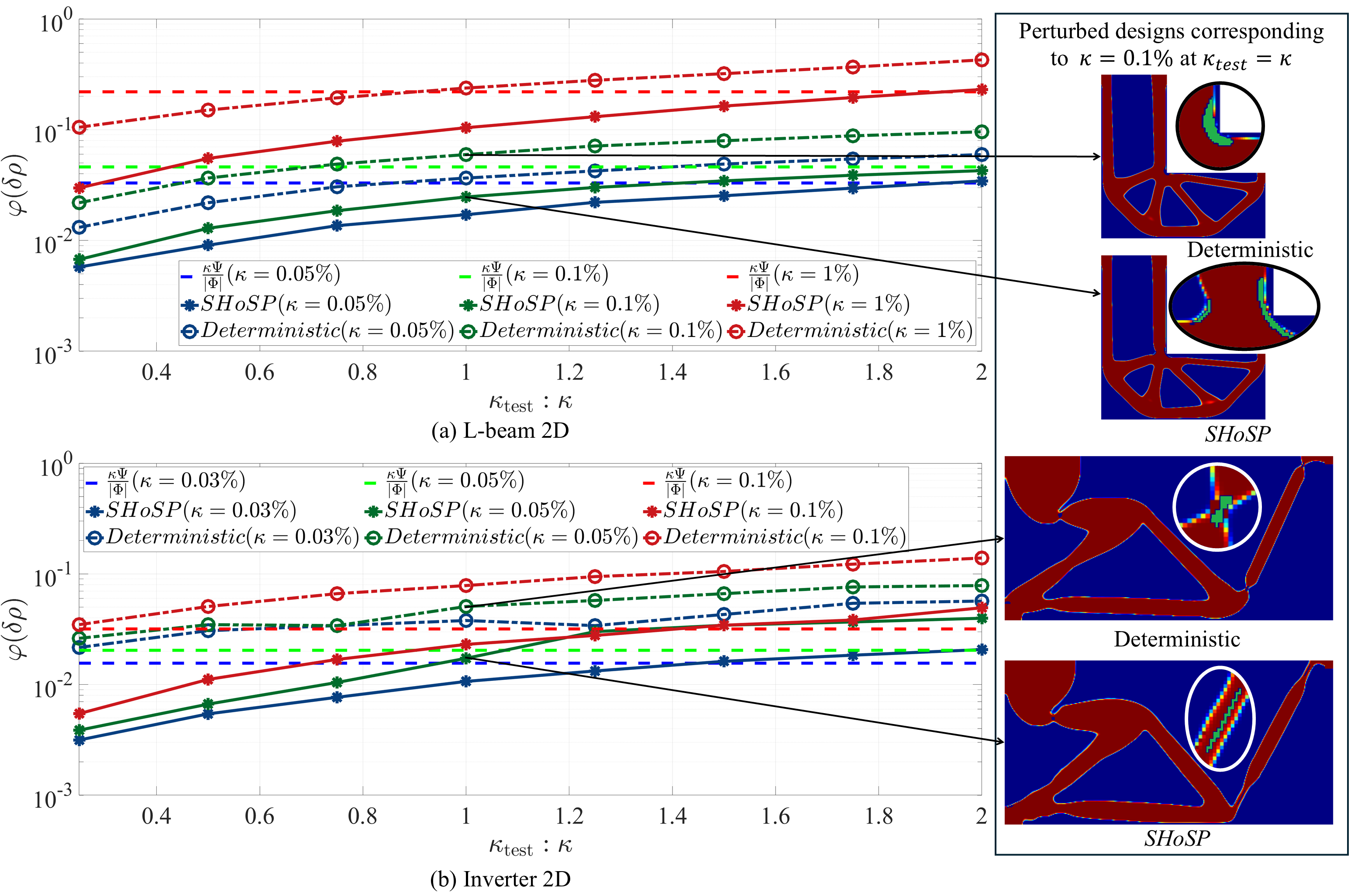}
    \caption{Perturbation-based robustness assessment using sensitivity-ranked adverse perturbations. The relative objective deterioration $\varphi$ is reported for the deterministic and SHoSP designs of the 2D L-beam and inverter benchmarks. For each SHoSP penalization level $\kappa$, the perturbation budget $\kappa_{\mathrm{test}}$ is varied relative to $\kappa$. The horizontal dashed lines indicate the first-order worst-case reference levels $\kappa\Psi/|\Phi|$ associated with the perturbation budgets considered in the corresponding SHoSP optimizations. Representative perturbed designs at $\kappa_{\mathrm{test}}=\kappa$ are shown on the right, where the regions marked in green denote the removed material.}
    \label{fig:RobustnessRanked}
\end{figure}

Figure~\ref{fig:RobustnessRanked} shows that the SHoSP designs exhibit consistently smaller objective deterioration than the deterministic designs for both benchmarks and for all penalization levels considered. More importantly, for every SHoSP design tested here, the measured deterioration remains below the corresponding first-order reference value $\kappa\Psi/|\Phi|$ whenever the testing budget satisfies $\kappa_{\mathrm{test}}\leq\kappa$. In contrast, the deterministic designs are not controlled by the SHoSP reference values and may exceed them, especially for larger values of $\kappa$ and $\kappa_{\mathrm{test}}$. The difference between deterministic and SHoSP responses increases with the testing budget, indicating that deterministic designs are more sensitive to concentrated adverse material-mass perturbations. These results provide direct numerical support for the interpretation that reducing the hot spot measure $\Psi$ lowers the sensitivity of the nominal objective to budgeted material-mass perturbations.

\begin{figure}
    \centering
    \includegraphics[width=1.0\linewidth]{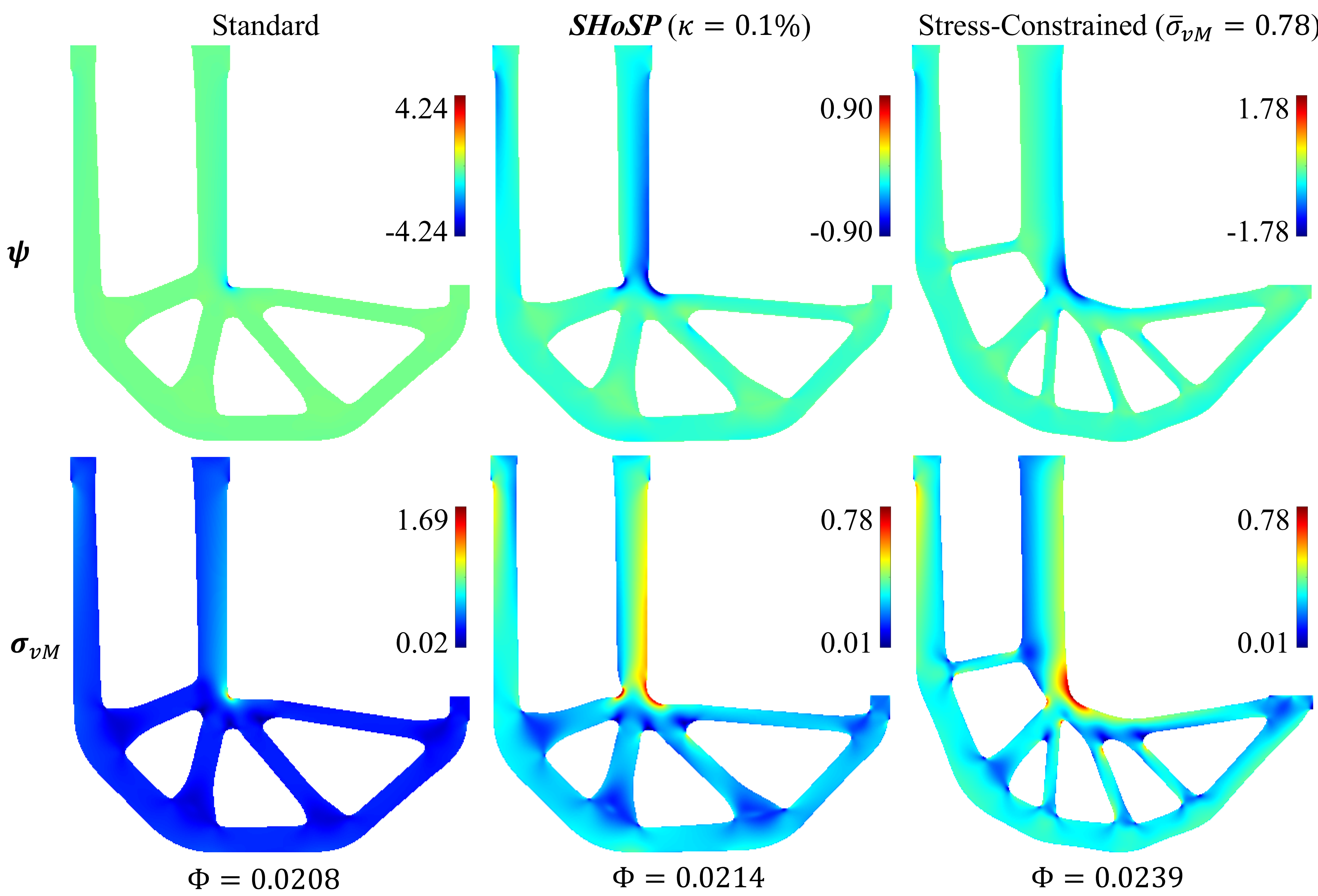}
    \caption{Comparison between the standard, SHoSP, and stress-constrained designs for the 2D L-beam benchmark. The rows show the sensitivity field $\bm{\psi}$ and the von Mises stress field $\bm{\sigma}_{vM}$, with the nominal compliance $\Phi$ reported below each design.}
    \label{fig:compar2StressTO_CoMi}
\end{figure}

Regarding the comparison with the conventional three-field robust formulation, it is not repeated here, since it was already carried out in the original SHoSP study~\cite{sigmund2026}. It is also worth emphasizing that the two methods correspond to different uncertainty models, i.e., norm-bounded material-mass perturbations and erosion/dilation-type geometric uncertainty, respectively. The robustness study above therefore focuses on validating the uncertainty interpretation established in Sec.~\ref{sec:NewInterp}.

\subsection{Comparison with stress-constrained topology optimization} \label{subsec:stressComparison}
Having established the characteristic SHoSP response on the two canonical benchmarks, we continue with the same examples to clarify how SHoSP differs from direct stress control. The numerical settings are kept identical to those in Subsec.~\ref{subsec:robustnessTest}, except that the stress-constrained formulation uses a $p$-norm aggregation of the von Mises stress with $p=32$. For each case, the stress bound $\bar{\sigma}_{vM}$ is chosen from the aggregated stress level of the corresponding SHoSP design, so that the comparison is made at a similar stress level. Following common practice in stress-constrained topology optimization, the maximum projection sharpness is limited to $\beta_{\max}=2R/(\sqrt{3}h_e)$ to reduce stress-evaluation errors near design boundaries~\cite{da2021three}. For compliance minimization, the stress constraint is active from the beginning of the optimization. For compliant mechanism design, it is activated after the first update of $\beta$, following the same delayed-activation strategy used for SHoSP. 

For the L-beam benchmark in Fig.~\ref{fig:compar2StressTO_CoMi}, both SHoSP and the stress-constrained formulation reduce the stress concentration near the re-entrant corner to approximately the same level. However, the resulting designs are not equivalent. SHoSP reduces the peak magnitude of $\bm{\psi}$ from $4.24$ to $0.90$, while increasing the compliance from $\Phi=0.0208$ to $\Phi=0.0214$. The stress-constrained design reaches a comparable stress level but exhibits a different topology and sensitivity distribution, with a higher compliance of $\Phi=0.0239$. The relatively high compliance of the stress-constrained result is also influenced by the adopted continuation strategy, which can lead to different local solutions for this problem. Accordingly, the comparison here is concerned with the resulting topology and sensitivity characteristics at comparable stress levels. Thus, even in compliance minimization, where sensitivity hot spots and stress concentrations are closely related, comparable stress reductions do not imply an equivalent sensitivity response or an optimized layout.

\begin{figure}
    \centering
    \includegraphics[width=1.0\linewidth]{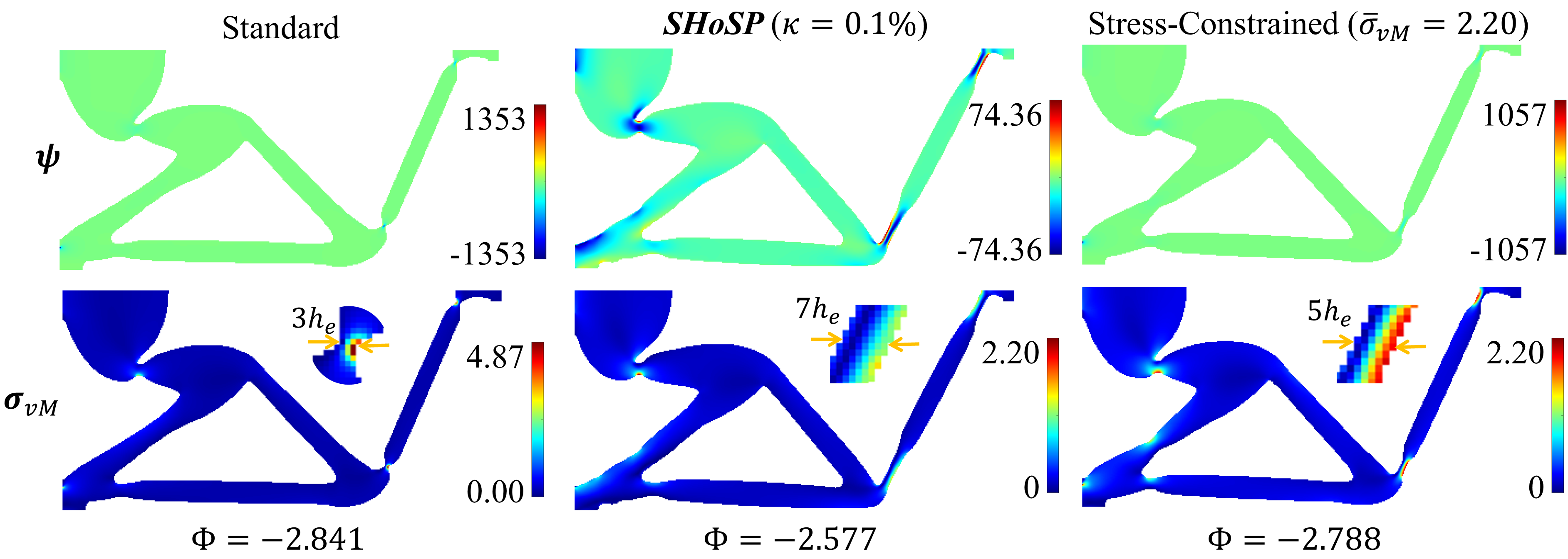}
    \caption{Comparison between the standard, SHoSP, and stress-constrained designs for the 2D inverter benchmark. The rows show the sensitivity field $\bm{\psi}$ and the von Mises stress field $\bm{\sigma}_{vM}$, with the nominal output objective $\Phi$ reported below each design.}
    \label{fig:compar2StressTO_CoMe}
\end{figure}

For the inverter benchmark in Fig.~\ref{fig:compar2StressTO_CoMe}, the distinction is more pronounced. The SHoSP and stress-constrained designs attain comparable maximum stress levels, but their sensitivity fields and resulting mechanisms differ markedly. SHoSP reduces the strong sensitivity localization of the deterministic design and replaces the localized hinge with a more distributed compliant region. This region is also visibly wider, with the characteristic width increasing from about $3h_e$ in the deterministic design to about $7h_e$ in the SHoSP design. The stress-constrained design forms a narrower compliant region of about $5h_e$ while satisfying the prescribed stress level. A pronounced sensitivity peak remains near the constrained region, contributing substantially to the reported maximum of $|\bm{\psi}|$. This localized value is therefore not used here as a direct measure of the relative performance of the two formulations. Rather, the comparison shows that direct control of the state-stress measure and suppression of objective-sensitivity localization can lead to different mechanisms even when similar stress levels are obtained. This distinction is consistent with the different quantities targeted by the two formulations: stress-constrained optimization acts directly on the state-stress response, whereas SHoSP suppresses the sensitivity concentration associated with the first-order worst-case response.

\subsection{Applicability to extended formulations} \label{subsec:extendedFormulations}
In this subsection, we examine the behavior of SHoSP in extended topology optimization formulations, including 3D problems, compliance minimization under multiple loading conditions, and porous infill optimization. These examples modify either the spatial dimension, the nominal objective aggregation, or the feasible design space, and are therefore used to assess how the characteristic SHoSP response and its sensitivity to the penalization parameter $\kappa$ change beyond the basic 2D setting. The corresponding histories of the relative SHoSP correction $r_\kappa=\kappa\Psi/|\Phi|$ are reported in Appendix~\ref{apdx:rkappa} as additional reference for interpreting the penalization levels used below.

\begin{figure}
    \centering
    \includegraphics[width=1.0\linewidth]{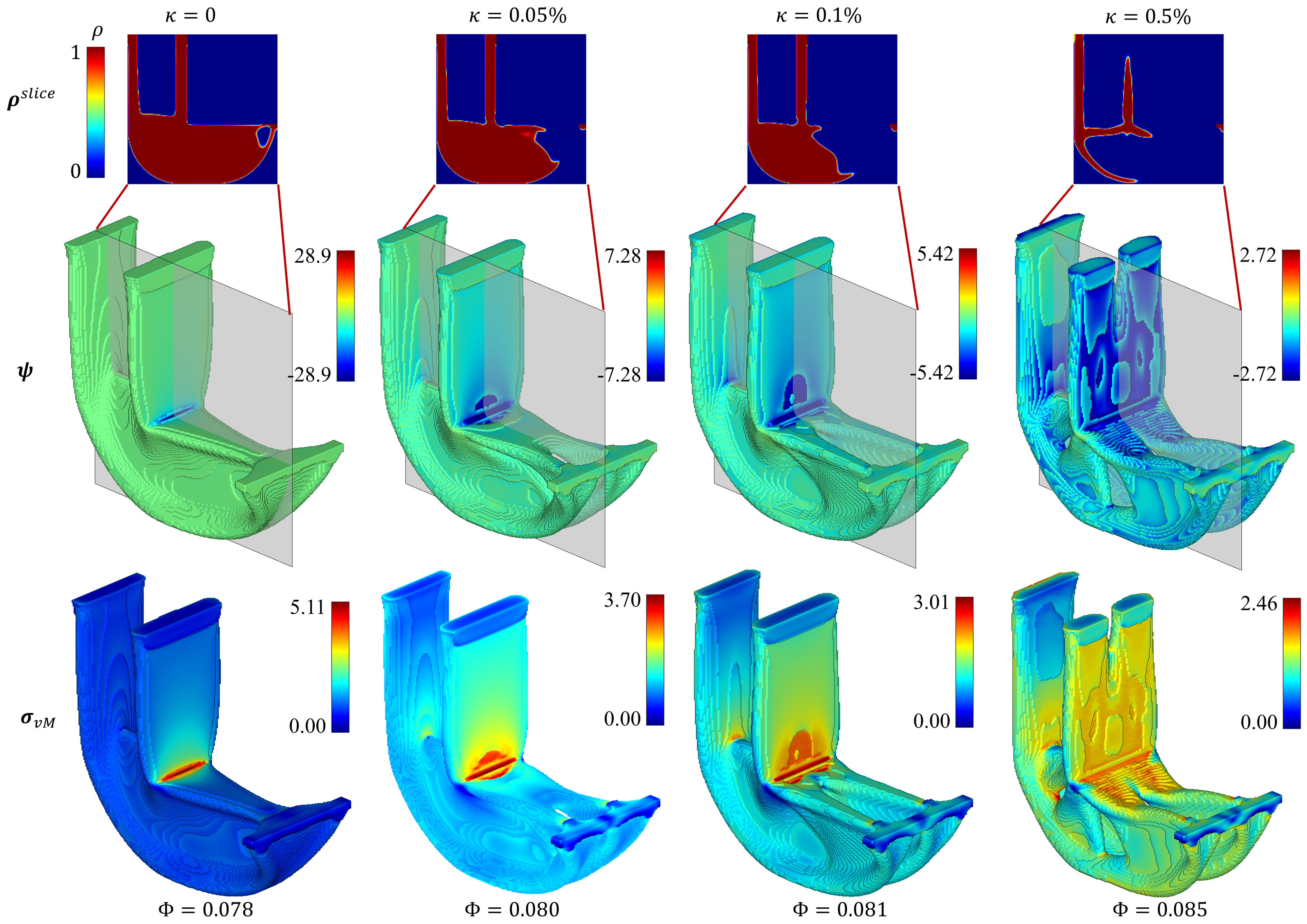}
    \caption{Effect of SHoSP on the 3D L-beam benchmark for different values of $\kappa$. The rows show density slices, the normalized sensitivity field $\bm{\psi}$, and the von Mises stress field $\bm{\sigma}_{vM}$ evaluated on the extracted Cartesian designs.}
    \label{fig:Lbeam3D}
\end{figure}

\paragraph{\textbf{3D inverter and L-beam benchmarks.}}
We first consider the 3D inverter and L-beam benchmarks shown in Fig.~\ref{fig:BenchmarkPDs}(c,d) to examine the SHoSP response and its dependence on $\kappa$ in genuinely 3D designs. Both examples are optimized with $R=12h_e$ and a prescribed volume fraction of $V_0=0.2$. The stress fields are directly evaluated on the structured meshes using the $\epsilon$-relaxation. For the L-beam, $\kappa=0$, $0.05\%$, $0.1\%$, and $0.5\%$ are considered, while for the inverter $\kappa=0$, $0.03\%$, $0.05\%$, and $0.1\%$ are used.

For the 3D L-beam in Fig.~\ref{fig:Lbeam3D}, increasing $\kappa$ strongly suppresses the sensitivity hot spot near the re-entrant region. The peak magnitude of $\bm{\psi}$ decreases from 28.9 in the deterministic design to 2.72 for $\kappa=0.5\%$. The corresponding designs exhibit a clear redistribution of material around the re-entrant region, as shown by the density slices. In this example, the reduction in sensitivity localization is also accompanied by a decrease in the maximum von Mises stress from $5.11$ to $2.46$. The nominal compliance increases only from $\Phi=0.078$ to $\Phi=0.085$, indicating that the reduction of sensitivity localization is obtained with a very moderate stiffness cost.

\begin{figure}
    \centering
    \includegraphics[width=1.0\linewidth]{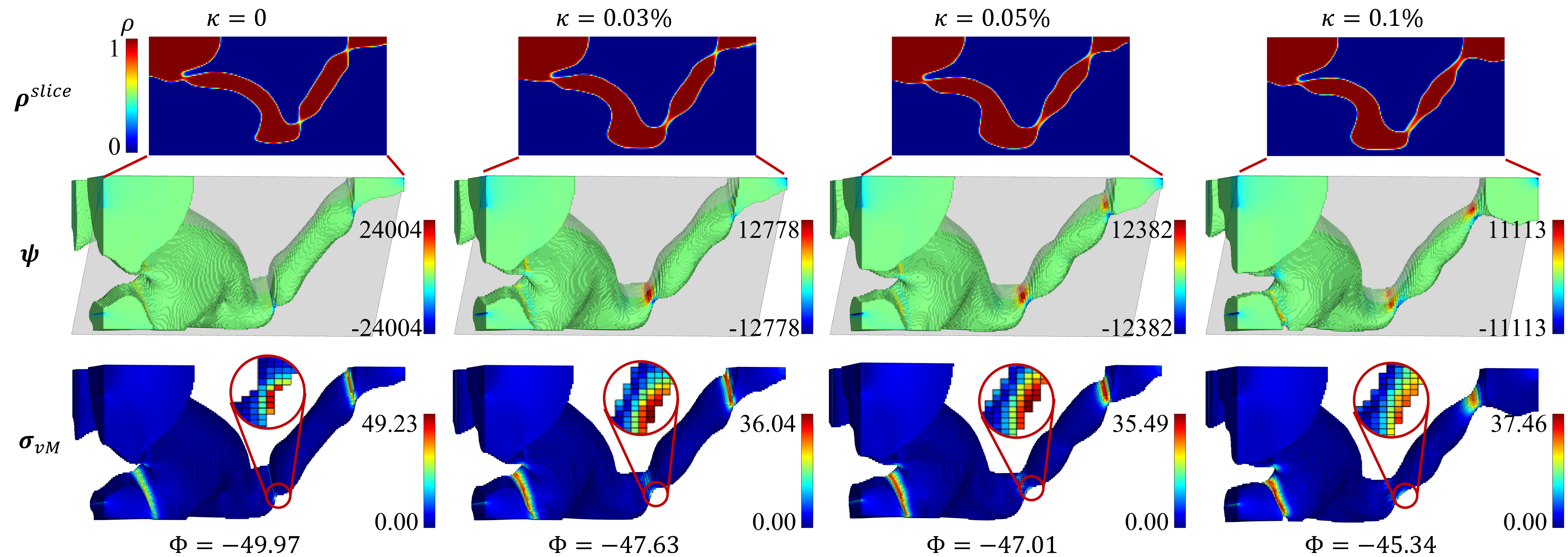}
    \caption{Effect of SHoSP on the 3D inverter benchmark for different values of $\kappa$. The results are organized using the same layout as in Fig.~\ref{fig:Lbeam3D}.}
    \label{fig:Inverter3D}
\end{figure}

The 3D inverter in Fig.~\ref{fig:Inverter3D} also shows reduced sensitivity localization. The peak magnitude of $\bm{\psi}$ decreases from $2.40\times10^4$ to $1.11\times10^4$, while the nominal output objective changes from $\Phi=-49.97$ to $\Phi=-45.34$. The stress field exhibits a redistribution of the local mechanical response, although small non-monotonic variations in the peak value are observed, since the $\epsilon$-relaxed stress measure remains sensitive to small differences in intermediate-density regions near highly stressed features. The compliant connection remains locally non-binary, particularly in the hinge region. This behavior is likely related to the filtering and projection setting: the adopted upper bound on $\beta$ was originally motivated by controlling the thickness of the projected transition layer for 2D solid-void interfaces, whereas a narrow 3D hinge is filtered simultaneously from multiple surrounding directions and may therefore retain intermediate densities even at the maximum projection sharpness. Nevertheless, the deterministic hinge-like feature becomes more spatially extended as $\kappa$ increases, indicating that the localized transmission mechanism is also weakened in 3D.

\begin{figure}
    \centering
    \includegraphics[width=1.0\linewidth]{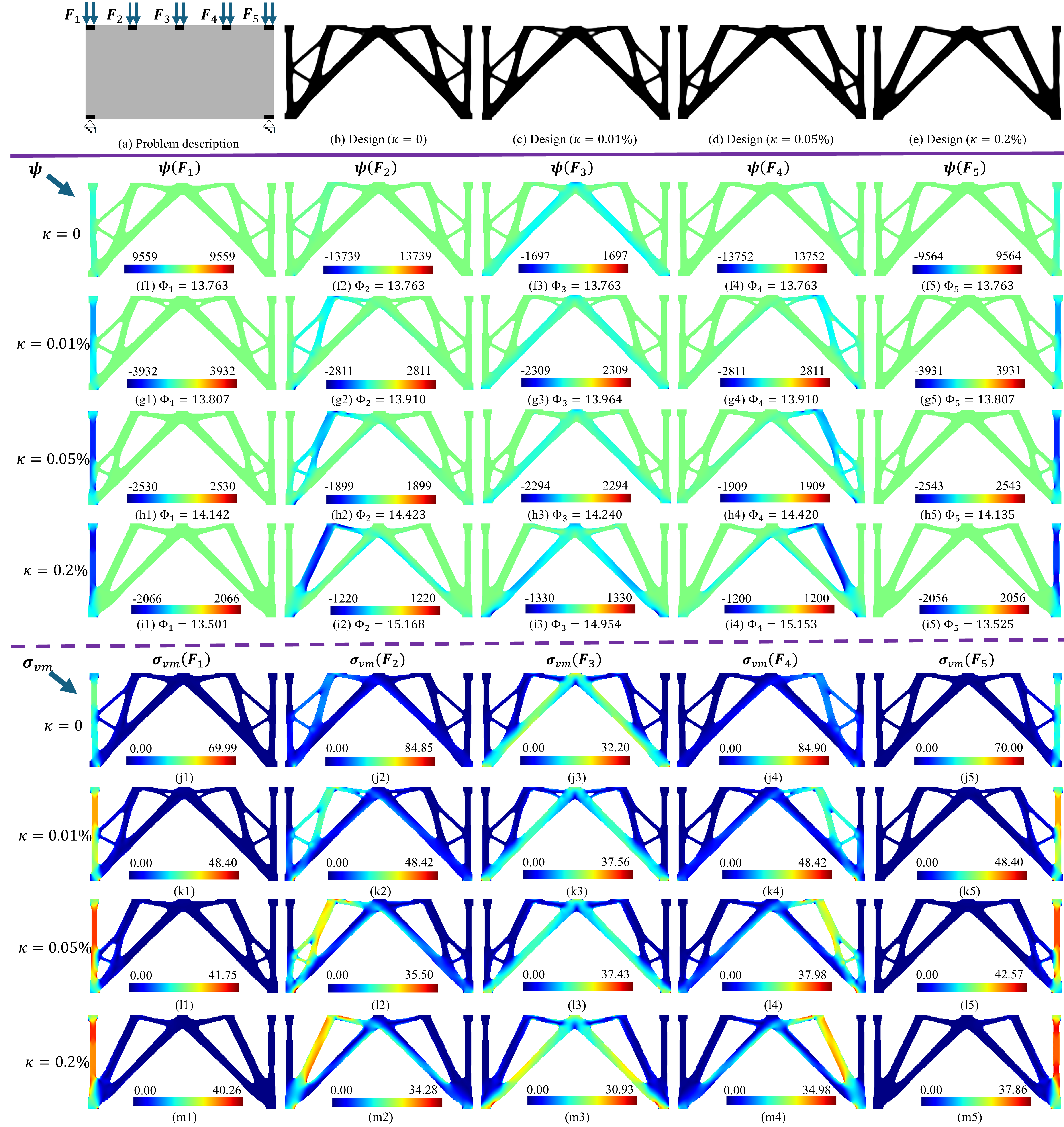}
    \caption{Multi-load bridge benchmark. (a) Problem description with five loading conditions. (b)-(e) Optimized designs for different values of $\kappa$. The following rows show the load-specific sensitivity fields $\bm{\psi}(F_i)$ together with the corresponding nominal compliances $\Phi_i$, followed by the von Mises stress fields $\bm{\sigma}_{vM}(F_i)$ for the individual load cases.}
    \label{fig:5loadBridge}
\end{figure}

\paragraph{\textbf{Multiple loading conditions.}}
We further examine the applicability of SHoSP to compliance minimization under multiple loading conditions using the 2D five-load bridge benchmark shown in Fig.~\ref{fig:5loadBridge}(a). The five load-specific SHoSP-augmented compliance responses are treated within the min--max formulation introduced in Sec.~\ref{subsec:extension}, while the global volume constraint remains unchanged. The filter radius is set to $R=24h_e$, and the prescribed volume fraction is $V_0=0.3$. We consider $\kappa=0$, $0.01\%$, $0.05\%$, and $0.2\%$.

As shown in Fig.~\ref{fig:5loadBridge}, increasing $\kappa$ suppresses the load-specific sensitivity hot spots and leads to a progressive redistribution of material among the principal load-carrying members. Across all five loading conditions, the largest sensitivity magnitude decreases from approximately $1.38\times10^4$ in the deterministic design to $2.07\times10^3$ for $\kappa=0.2\%$. At the same time, the nominal compliances become more differentiated among the individual load cases, since the min--max formulation balances the SHoSP-augmented responses $\Phi_i+\kappa\Psi_i$ rather than the nominal compliances $\Phi_i$ alone. For $\kappa=0.2\%$, the largest nominal compliance is $\Phi_2=15.168$, compared with $\Phi_i=13.763$ for the deterministic design. The corresponding convergence histories for $\kappa=0.2\%$ are provided in Appendix~\ref{apdx:multiConvergence} for reference.

The stress fields exhibit a corresponding redistribution across the individual loading conditions. The pronounced stress concentrations associated with the off-center load cases $F_2$ and $F_4$ are substantially reduced, with their peak von Mises stresses decreasing from approximately $84.9$ in the deterministic design to $34.3$ and $35.0$, respectively, for $\kappa=0.2\%$. The response under the central load $F_3$ changes less markedly, while the outer load cases $F_1$ and $F_5$ also exhibit clear reductions in peak stress. These results show that treating the load-specific SHoSP responses within a min--max formulation suppresses sensitivity localization across the competing loading conditions without requiring prescribed weighting factors. They also demonstrate that SHoSP-augmented responses can be used effectively within optimization constraints, extending the method beyond a single augmented objective.

\paragraph{\textbf{Porous infill optimization.}}
We next consider the 2D L-beam benchmark with a local volume constraint to examine the applicability of SHoSP to porous infill optimization. 
The porosity is controlled by setting $R_e=15h_e$ and $V_{e0}=0.6$, see Eq.~\ref{eqn:LVF}, while an additional global volume constraint of $V_0=0.5$ fixes the total material usage. To obtain a clearly resolved porous structure, the maximum projection sharpness is set to $\beta_{\max}=128$, following Refs.~\cite{wu2017infill,wang2022stress}, with the continuation starting from $\beta=1$ and doubling every 50 optimization steps. The filter radius is set to $R=4h_e$. Four penalization intensities are considered: $\kappa=0$, $0.03\%$, $0.05\%$, and $0.10\%$.

\begin{figure}
    \centering
    \includegraphics[width=1.0\linewidth]{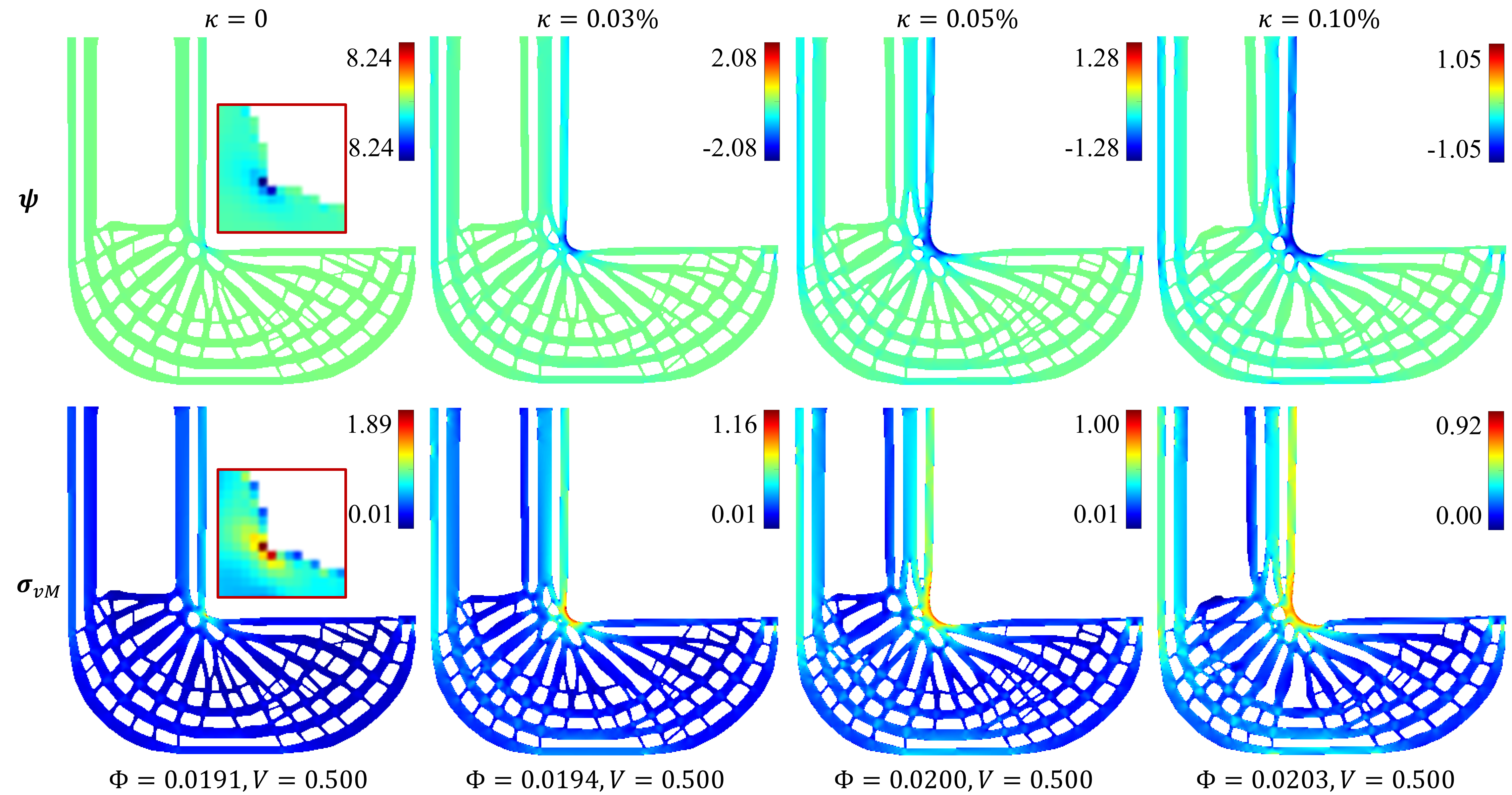}
    \caption{Porous infill optimization of the 2D L-beam with local and global volume constraints. The rows show the sensitivity field $\bm{\psi}$ and the $\epsilon$-relaxed von Mises stress field $\bm{\sigma}_{vM}$ for different values of $\kappa$. The nominal compliance $\Phi$ and final volume fraction $V$ are reported below each design.}
    \label{fig:2DPIO}
\end{figure}

As shown in Fig.~\ref{fig:2DPIO}, SHoSP retains its characteristic hot-spot suppression behavior despite the restricted material redistribution imposed by the local volume constraints. As $\kappa$ increases, the sensitivity concentration near the re-entrant corner is progressively reduced, accompanied by a redistribution of the surrounding porous members. The peak magnitude of $\bm{\psi}$ decreases from $8.24$ in the deterministic design to $1.05$ for $\kappa=0.10\%$, while the maximum von Mises stress decreases from $1.89$ to $0.92$. Over the same range, the nominal compliance increases from $\Phi=0.0191$ to $\Phi=0.0203$, while the prescribed global volume fraction $V=0.5$ is maintained. These results show that the SHoSP mechanism remains effective when the admissible material redistribution is further restricted by local volume constraints, with substantial reductions in sensitivity localization accompanied by a modest increase in compliance.

\begin{figure}
    \centering
    \includegraphics[width=1.0\linewidth]{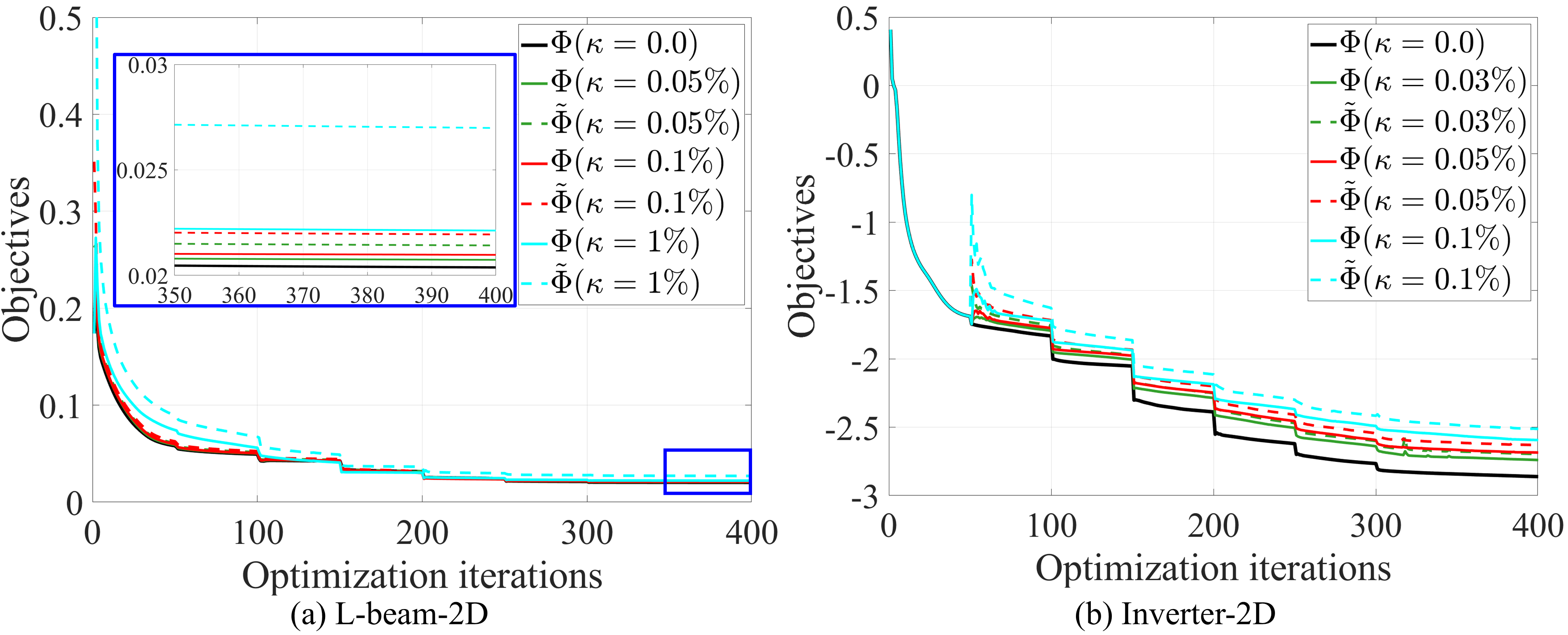}
    \caption{Convergence histories of the nominal objective $\Phi$ and the augmented objective $\tilde{\Phi}$ for the 2D L-beam and inverter benchmarks.}
    \label{fig:convergence}
\end{figure}

\paragraph{\textbf{Convergence behavior and computational overhead.}}
We finally examine the numerical behavior of SHoSP in terms of convergence and computational overhead. Figure~\ref{fig:convergence} shows the optimization histories of the nominal objective $\Phi$ and the augmented objective $\tilde{\Phi}$ for the 2D L-beam and inverter benchmarks. For the L-beam, the penalized cases retain the smooth convergence behavior of standard compliance minimization, with a moderate separation between $\Phi$ and $\tilde{\Phi}$ that increases with $\kappa$. For the inverter, the histories exhibit jumps associated with the projection continuation, but the penalty term does not introduce irregular or unstable convergence. Thus, for the penalization levels considered here, SHoSP changes the optimized designs without degrading the basic convergence behavior of the underlying topology optimization problem.

\begin{figure}
    \centering
    \includegraphics[width=1.0\linewidth]{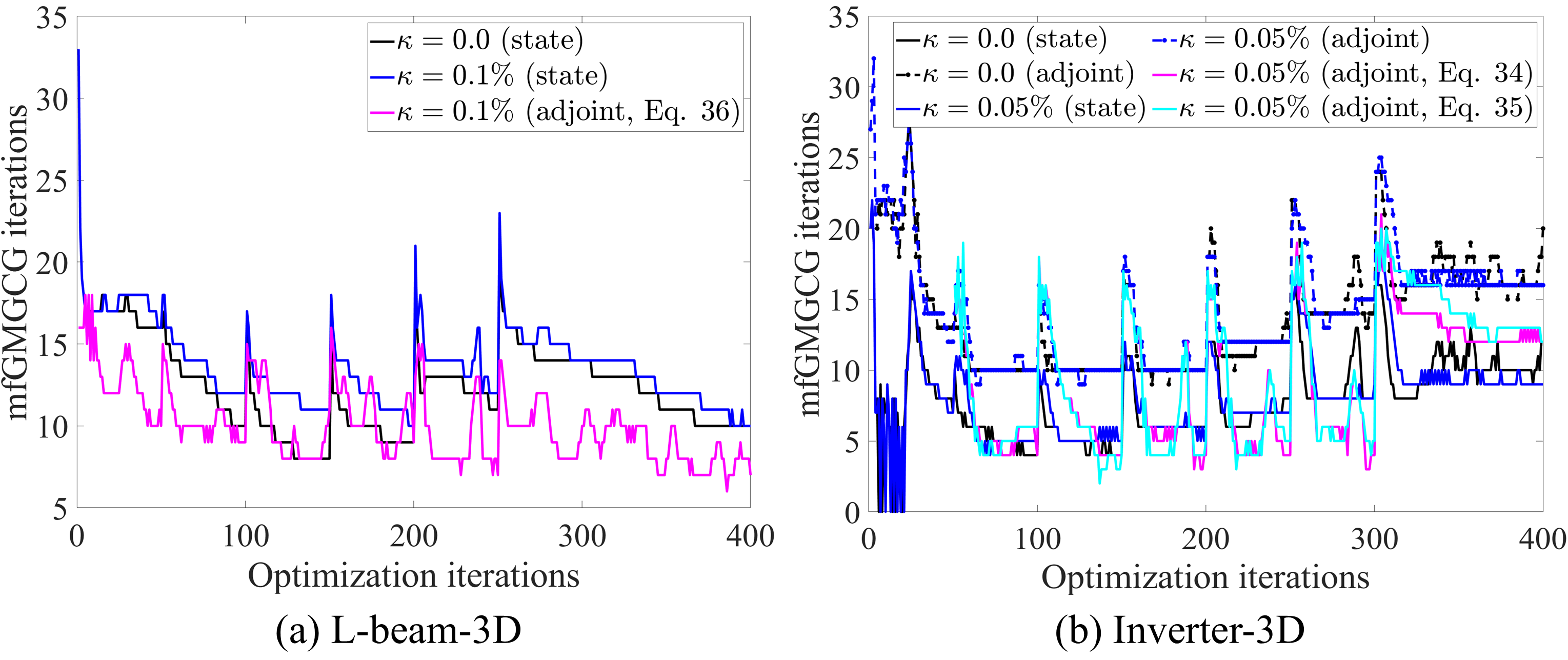}
    \caption{Numbers of mfGMGCG iterations required by the linear systems solved during the 3D optimization histories. The L-beam results compare the state equation and the SHoSP adjoint system. The inverter results compare the state equation, the nominal adjoint equation, and the two SHoSP adjoint systems.}
    \label{fig:solvingIts}
\end{figure}

The additional computational effort is assessed using the 3D L-beam and inverter benchmarks above. Figure~\ref{fig:solvingIts} reports the numbers of mfGMGCG iterations required by the state, nominal adjoint (for inverter), and SHoSP adjoint systems during optimization. For the 3D L-beam, the SHoSP adjoint system requires a number of iterations comparable to, and often lower than, the corresponding state system. A similar trend is observed for the 3D inverter: the two SHoSP adjoint systems remain comparable to the state system, whereas the nominal adjoint system can be more demanding, especially in the later iterations. These results support the scalability discussion in Sec.~\ref{subsec:cmpt}: although SHoSP introduces additional RHSs, their iterative solution difficulty remains comparable to that of the original state and adjoint equations. The iteration histories also confirm that the SHoSP auxiliary systems remain well suited for warm starts despite their iteration-dependent right-hand sides.

\subsection{Large-scale practical demonstrations} \label{subsec:largeScale}
The preceding examples verified the behavior of SHoSP on controlled benchmark problems and extended optimization settings. We now demonstrate its applicability to larger and more practical 3D design problems. A single representative penalization level, $\kappa=0.05\%$, is used for all examples. Three additional 3D cases are considered: a compliant gripper for compliant mechanism design, the GE bracket under multiple loading conditions, and a molar structure for porous infill optimization. The computational sizes and wall-clock times are summarized in Table~\ref{tab:models}.

\begin{table*}[ht]
\centering
\caption{Computational sizes and wall-clock times of the 3D examples, incl. the two examples covered in Sec.~\ref{subsec:extendedFormulations}.}
\begin{tabular}{c|c|r|r|r|r}
\hline
\hline
Models                      & $\#$Resolution                            & $\#$Elements                  & $\#$DOFs                      & $\kappa$  & Timings: hours \\
\hline
\multirow{4}{*}{L-beam}     & \multirow{4}{*}{$200\times100\times200$}  & \multirow{4}{*}{$2{,}560{,}000$}    &  \multirow{4}{*}{$7{,}878{,}303$}   &  $0$              &  $0.73$            \\
\cline{5-6}
                            &                                           &                               &                               &  $0.05\%$              &  $1.00$        \\
\cline{5-6}
                            &                                           &                               &                               &  $0.10\%$              &  $1.01$        \\   
\cline{5-6}
                            &                                           &                               &                               &  $0.50\%$              &  $1.13$        \\                           
\cline{5-6}
\hline
\multirow{4}{*}{Inverter}   & \multirow{4}{*}{$200\times100\times100$}  & \multirow{4}{*}{$2{,}000{,}000$}    &  \multirow{4}{*}{$6{,}151{,}203$}   &  $0$              &  $1.12$           \\
\cline{5-6}
                            &                                           &                               &                               &  $0.03\%$              &  $1.54$        \\
\cline{5-6}
                            &                                           &                               &                               &  $0.05\%$              &  $1.50$        \\   
\cline{5-6}
                            &                                           &                               &                               &  $0.10\%$              &  $1.50$        \\                               
\hline                    
\multirow{2}{*}{Gripper}    & \multirow{2}{*}{$600\times300\times300$}  & \multirow{2}{*}{$54{,}000{,}000$}   &  \multirow{2}{*}{$155{,}253{,}603$} &  $0$              &  $29.44$            \\
\cline{5-6}
                            &                                           &                               &                               &  $0.05\%$              &  $43.06$        \\
\hline
\multirow{2}{*}{GE Bracket} & \multirow{2}{*}{$311\times512\times180$}  & \multirow{2}{*}{$10{,}991{,}705$}   &  \multirow{2}{*}{$33{,}752{,}853$}  &  $0$              &  $7.12$            \\
\cline{5-6}
                            &                                           &                               &                               &  $0.05\%$              &  $11.93$        \\
\hline
\multirow{2}{*}{Molar}      & \multirow{2}{*}{$700\times431\times463$}  & \multirow{2}{*}{$43{,}638{,}295$}   &  \multirow{2}{*}{$132{,}743{,}505$} &  $0$              &  $36.64$           \\
\cline{5-6}
                            &                                           &                               &                               &  $0.05\%$              &  $65.86$        \\
\hline
\hline
\end{tabular}
\label{tab:models}
\end{table*}

\begin{figure}
    \centering
    \includegraphics[width=1.0\linewidth]{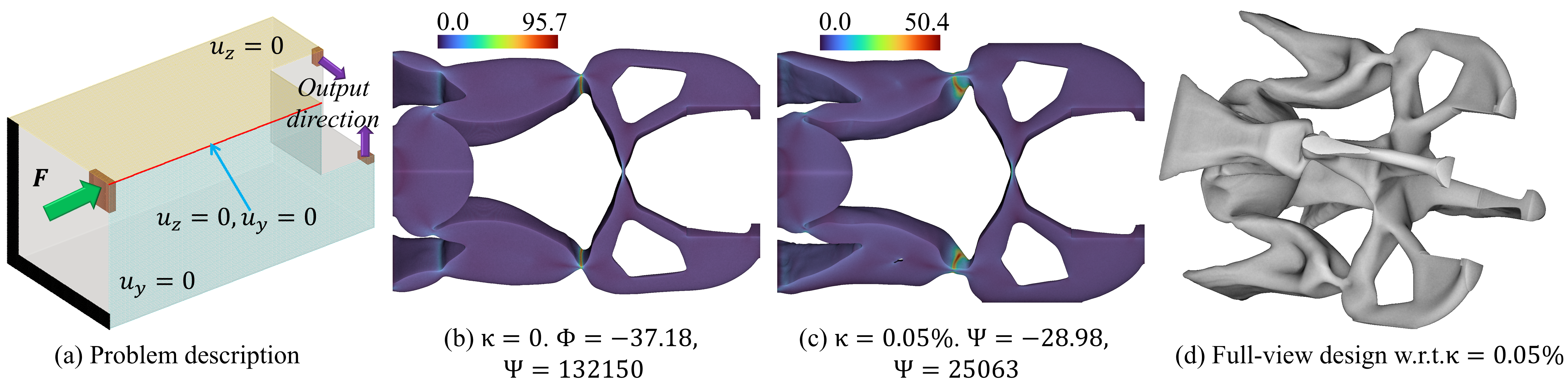}
    \caption{Large-scale 3D compliant gripper. (a) Quarter-domain problem description. (b,c) Optimized designs and von Mises stress fields for the deterministic and SHoSP formulations. (d) Full-view reconstruction of the SHoSP design.}
    \label{fig:Gripper}
\end{figure}

\paragraph{\textbf{Compliant gripper.}}
We first consider a compliant gripper whose full design domain is symmetric with respect to two planes and has an equivalent resolution of $600\times600\times600$. Figure~\ref{fig:Gripper}(a) shows the corresponding quarter-domain model used in the computation, analogous to Fig.~\ref{fig:BenchmarkPDs}(c). The filter radius is set to $R=36h_e$, and the prescribed volume fraction is $V_0=0.1$.

Figures~\ref{fig:Gripper}(b,c) compare the deterministic and SHoSP designs. The deterministic design relies on highly localized hinge-like connections, whereas the SHoSP design replaces them with thicker and more spatially extended compliant regions. The aggregated hot-spot measure is reduced from $\Psi=1.32\times10^5$ to $\Psi=2.51\times10^4$, and the maximum von Mises stress decreases from $95.7$ to $50.4$. This is obtained with a reduced nominal output performance, with $\Phi$ changing from $-37.18$ to $-28.98$. The full reconstructed SHoSP design is shown in Fig.~\ref{fig:Gripper}(d). 

\begin{figure}
    \centering
    \includegraphics[width=1.0\linewidth]{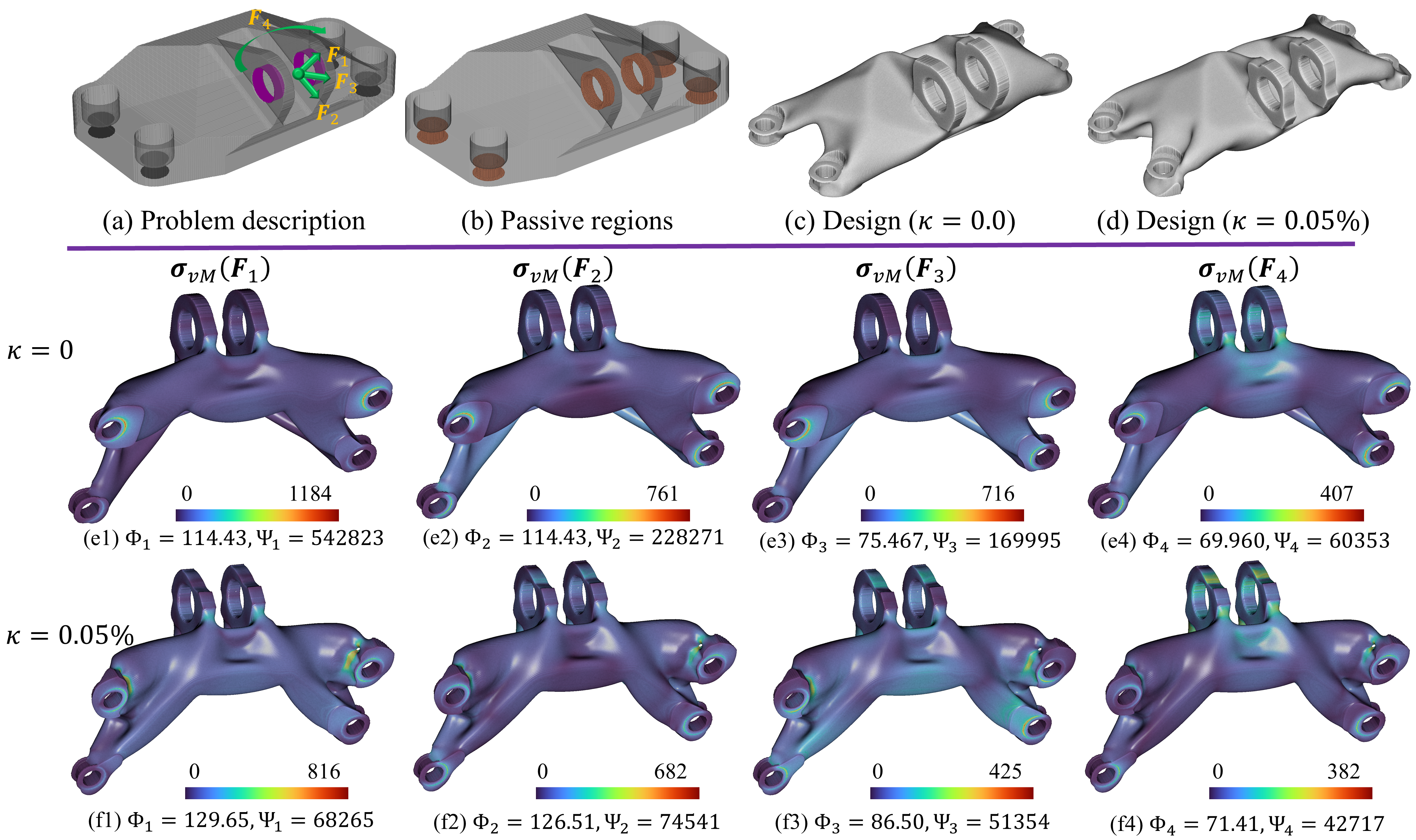}
    \caption{Large-scale GE bracket under multiple loading conditions. (a) Problem description with four load cases. (b) Prescribed passive regions. (c,d) Deterministic and SHoSP designs. (e1)-(e4) and (f1)-(f4) show the corresponding von Mises stress fields for the individual load cases, together with the compliance $\Phi_i$ and sensitivity hot spot measure $\Psi_i$.}
    \label{fig:GEBracket}
\end{figure}

\paragraph{\textbf{GE bracket.}}
The second example is the GE bracket, used here for compliance minimization under multiple loading conditions. Four load cases are considered, as shown in Fig.~\ref{fig:GEBracket}(a), where the torque load case $\bm{F}_4$ is approximated by a pair of opposite forces applied on the inner surfaces of the two loaded lugs. The prescribed passive regions are shown in Fig.~\ref{fig:GEBracket}(b). The optimization is performed with $R=24h_e$ and $V_0=0.3$.

Figures~\ref{fig:GEBracket}(c,d) compare the deterministic and SHoSP designs. For the deterministic design, the compliances of the four load cases are $[114.43,114.43,75.47,69.96]$, with the first two load cases governing the min--max objective. For the SHoSP design with $\kappa=0.05\%$, the corresponding compliance values become $[129.65,126.51,86.50,71.41]$. The first two load cases also remain governing for the SHoSP design, while the sensitivity hot spot measures decrease from $[5.43\times10^5,2.28\times10^5,1.70\times10^5,6.04\times10^4]$ to $[6.83\times10^4,7.45\times10^4,5.14\times10^4,4.27\times10^4]$, respectively. The stress fields in Fig.~\ref{fig:GEBracket}(e,f) show a corresponding redistribution of the local geometry. The maximum von Mises stresses for $\bm F_1$-$\bm F_4$ decrease from $1184$, $761$, $716$, and $407$ to $816$, $682$, $425$, and $382$, respectively. Overall, the example demonstrates substantial hot-spot reduction across the individual load cases in a large-scale multi-load bracket design, with a moderate stiffness trade-off.

\begin{figure}
    \centering
    \includegraphics[width=1.0\linewidth]{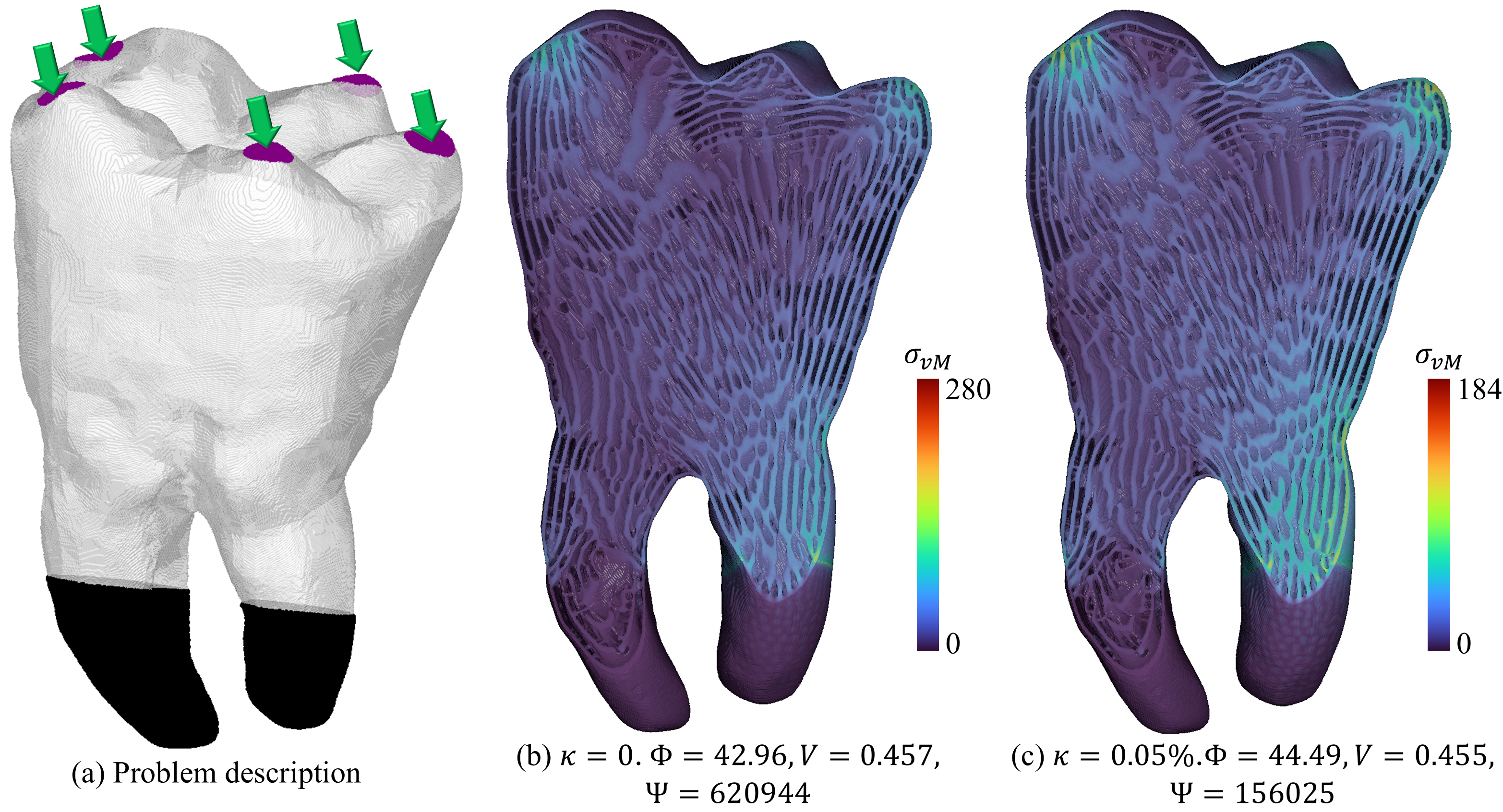}
    \caption{Large-scale porous infill optimization of a molar structure. (a) Problem description with prescribed loading and fixed root regions. (b,c) Deterministic and SHoSP porous infill designs with the corresponding von Mises stress fields. Cut views are used to expose the internal porous layouts.}
    \label{fig:Molar}
\end{figure}

\paragraph{\textbf{Molar.}}
The final example considers porous infill optimization of the molar structure shown in Fig.~\ref{fig:Molar}(a). A local volume constraint is imposed with $V_{e0}=0.5$ and $R_e=6h_e$, while the filter radius is set to $R=2h_e$. The maximum projection sharpness is set to $\beta_{\max}=128$ to obtain a clearly resolved porous structure.

Figures~\ref{fig:Molar}(b,c) compare the deterministic and SHoSP porous infill designs. The global geometric differences are modest because both designs are constrained to form distributed porous layouts, but the quantitative improvement is clear. The aggregated hot spot measure decreases from $\Psi=6.21\times10^5$ to $\Psi=1.56\times10^5$, and the maximum von Mises stress decreases from $280$ to $184$. The compliance increases only from $\Phi=42.96$ to $\Phi=44.49$, while the final volume fraction remains essentially unchanged. The larger relative overhead compared with the gripper example is consistent with the highly heterogeneous stiffness distribution of porous infill designs, which increases the cost of each linear solve. Overall, the example demonstrates that SHoSP remains effective for large-scale porous infill optimization, even when material redistribution is strongly restricted by local volume constraints.

\section{Conclusion and future work}\label{sec:conclusion}
This work takes a deeper dive into sensitivity hot spot penalization (SHoSP) in density-based topology optimization, revealing its underlying interpretation as a robust topology optimization framework against first-order worst-case perturbations.
By introducing element-wise material-mass perturbations and applying a first-order Taylor expansion together with H\"older's duality, the SHoSP-augmented objective was shown to correspond to a worst-case first-order approximation under a budgeted material-mass uncertainty set. In this formulation, the penalty parameter $\kappa$ has a direct meaning as a dimensionless norm-bounded material-mass perturbation budget relative to the design-domain area or volume.
The large-$p$ aggregation used in SHoSP corresponds to a near-$\ell_1$ material-mass uncertainty mode, which is consistent with the intended suppression of localized sensitivity hot spots. The perturbation-based robustness assessment further supports this connection by showing that SHoSP designs exhibit smaller objective deterioration under density-bounded adverse material-mass perturbations.

The mechanical implications of sensitivity hot spots have also been clarified. For compliance minimization, sensitivity hot spots are closely related to density-weighted local stress-energy concentrations, which explains why suppressing them often alleviates stress concentrations. For compliant mechanism design, the corresponding hot spots arise from the local interaction between the state and adjoint stress fields, and therefore identify localized force-transmission mechanisms rather than peaks in a single stress field. This distinction also clarifies why SHoSP and stress-constrained topology optimization can lead to different designs, even when similar stress levels are obtained. 
The same formulation was further shown to remain applicable to extended topology optimization settings, including multiple loading conditions and porous infill optimization, where SHoSP-augmented responses can enter directly as constraint functions in the min--max formulation, while the local volume constraints restrict the admissible material redistribution.

The numerical examples confirm that SHoSP can substantially reduce the aggregated hot-spot measure $\Psi$ with moderate changes in the nominal objective $\Phi$ in both 2D and 3D settings. Large-scale examples demonstrate that the method remains computationally feasible at scale. Although SHoSP introduces additional adjoint systems, these systems share the same stiffness matrix as the original analysis, and the evolving auxiliary right-hand sides remain compatible with warm-started iterative solves in the examples considered.

The first-order worst-case approximation framework developed here suggests several natural extensions. One direction is to explore other perturbation modes within the same material-mass uncertainty setting. The present work retains the original SHoSP choice of a large aggregation exponent ($p=32$), which corresponds to a near-$\ell_1$ perturbation budget and is well aligned with the suppression of localized hot spots. Other choices of $p$ would correspond to different admissible perturbation geometries and may be useful for different goals. For example, an $\ell_2$-type uncertainty mode may favor the reduction of a more distributed sensitivity measure rather than isolated peaks. More general sensitivity aggregations or transformations may also be considered for numerical regularization, although their associated uncertainty interpretations would need to be derived separately. Another direction is to introduce application-specific uncertainty descriptions, including spatially correlated errors in material content, directional over- or under-realization of material, and manufacturing-process-dependent perturbation fields. These models can be incorporated by modifying the perturbation norm, the admissible perturbation space, or the sensitivity aggregation used in the augmented objective. For applications in which the admissible perturbations are no longer small, higher-order or finite-amplitude extensions of the present approximation may also be relevant. Further work may also consider adaptive strategies for choosing $\kappa$ during continuation and the integration of SHoSP into open large-scale topology optimization tools.

\section*{Declaration of competing interest}
The authors declare that they have no known competing financial interests or personal relationships that could have appeared to influence the work reported in this paper.

\section*{Data availability}
Data will be made available on request.

\section*{Declaration of generative AI and AI-assisted technologies in the manuscript preparation process}
During the preparation of this work, the authors used ChatGPT (OpenAI) and Claude (Anthropic) for language editing, proofreading, and improving the clarity of the manuscript. The authors reviewed and edited the output as needed and take full responsibility for the content of the published article.

\printcredits

\section*{Acknowledgment}
We acknowledge the financial support from the Villum Foundation through the Villum Investigator Project Amstrad (VIL54487).

\appendix
\section{Robin boundary parameter in the PDE filter}\label{apdx:RobinBC}

The PDE filter used in this work includes the Robin boundary treatment of Wallin et al.~\cite{wallin2020consistent}, which compensates for missing neighborhood information outside non-box-shaped design domains. The compensation strength is controlled by the parameter $l_s$ in Eq.~\ref{eqn:PDEfilter}. Since this parameter mainly affects the filtering behavior near free design-domain boundaries, we include a short numerical check to justify the value used in the present computations.

\begin{figure}
    \centering
    \includegraphics[width=1.0\linewidth]{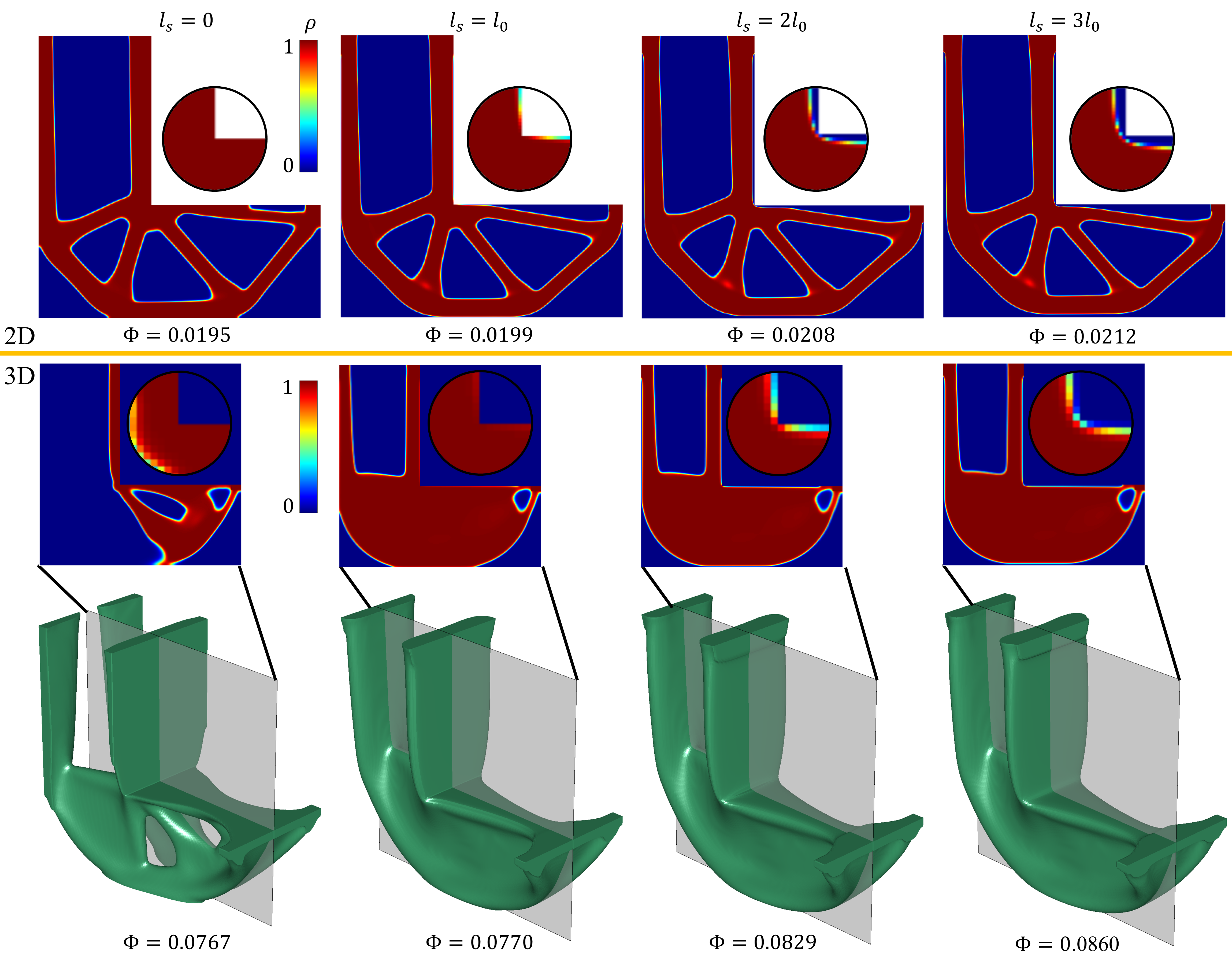}
    \caption{Effect of the Robin boundary parameter $l_s$ in the PDE filter for the 2D and 3D L-beam benchmarks. The value $l_s=2l_0$ is used throughout the paper as a practical compromise between boundary compensation and excessive rounding near design-domain creases.}
    \label{fig:RobinBC}
\end{figure}

Figure~\ref{fig:RobinBC} shows deterministic compliance minimization results for the 2D and 3D L-beam benchmarks using $l_s=0$, $l_0$, $2l_0$, and $3l_0$, while keeping all other parameters unchanged. Without the Robin compensation, material tends to accumulate near the free boundaries due to the missing exterior filtering neighborhood. Increasing $l_s$ progressively compensates for this boundary effect and produces smoother boundary behavior, but overly strong compensation can round the design-domain creases and slightly increase the objective value. Based on this comparison, we use $l_s=2l_0$ throughout the paper as a practical compromise. More details about Robin boundary condition for PDE filters can be found in~\cite{wallin2020consistent}.

\section{Derivation of $\frac{\mathrm{d}\Psi} {\mathrm{d}\rho_e}$} \label{apdx:Adjoint}
This appendix provides a compact derivation of the sensitivity of the hot spot measure, i.e., the total derivative of $\Psi$ with respect to the density variable ($\rho_e$), which leads to the auxiliary adjoint systems used in Sec.~\ref{subsec:adjoint}. The derivation is equivalent to the adjoint-variable procedure used in the original SHoSP~\cite{sigmund2026}, but is written here in a compact direct-differentiation form that starts from the sensitivity-density aggregation adopted in this work.

Since compliance minimization is self-adjoint, it can be regarded as a simplified special case of the compliant mechanism formulation under the sign convention adopted in this work. We therefore first present the derivation for compliant mechanism design, and then state how the compliance result follows from the self-adjoint relation. 

Recall Eqs.~\ref{eqn:MechanismSensitivity}, \ref{eqn:augObjSensDef}, \ref{eqn:penaSens}, and \ref{eqn:wDefi}. The critical step is to determine the derivative of the sensitivity-density field with respect to the physical density variables. The derivative of the hot spot measure is written as
\begin{equation} \label{eqn:psi2psistar}
    \frac{\mathrm{d}\Psi}{\mathrm{d}\rho_e} = \sum_i w_i \frac{\mathrm{d}\psi_i}{\mathrm{d}\rho_e},
\end{equation}
The remaining task is therefore to evaluate $\frac{\mathrm{d}\psi_i}{\mathrm{d}\rho_e}$. For compliant mechanism design, $\psi_i$ depends on $\bm{\rho}$ explicitly and through the state and nominal adjoint fields ($\bm{U}$ and $\bm{\lambda}$) implicitly. Its total derivative with respect to $\rho_e$ is 
\begin{equation} \label{eqn:fullDeri}
    \frac{\mathrm{d}\psi_i}{\mathrm{d}\rho_e}
    =
    \frac{\partial \psi_i}{\partial \rho_e}
    +
    \frac{\partial \psi_i}{\partial \bm{U}}
    \frac{\mathrm{d} \bm{U}}{\mathrm{d} \rho_e}
    +
    \frac{\partial \psi_i}{\partial \bm{\lambda}}
    \frac{\mathrm{d} \bm{\lambda}}{\mathrm{d} \rho_e}.
\end{equation}
Differentiating the state and nominal adjoint equations (cf. Eqs.~\ref{eqn:FEA} and \ref{eqn:MechanismAdjoint}) with respect to $\rho_e$, and rearranging gives
\begin{equation}\label{eqn:differentiateStateAdjoint}
    \frac{\mathrm{d} \bm{U}}{\mathrm{d} \rho_e} = -\bm{K}^{-1}\frac{\partial \bm{K}}{\partial \rho_e}\bm{U}, 
    \qquad 
    \frac{\mathrm{d} \bm{\lambda}}{\mathrm{d} \rho_e} = -\bm{K}^{-1}\frac{\partial \bm{K}}{\partial \rho_e}\bm{\lambda}
\end{equation}
According to the definition of $\psi_i$, and leveraging the symmetry of $\bm{K}$, one obtains
\begin{equation}\label{eqn:chainRule4PartialPsiEleCompact}
    \frac{\partial \psi_i}{\partial\bm{U}} = \left(\frac{A}{A_i} \frac{\partial \bm{K}}{\partial \rho_i}\bm{\lambda} \right)^T, 
    \qquad 
    \frac{\partial \psi_i}{\partial\bm{\lambda}} = \left(\frac{A}{A_i} \frac{\partial \bm{K}}{\partial \rho_i}\bm{U} \right)^T
\end{equation}
Since each element stiffness depends only on its own density, the explicit term in Eq.~\ref{eqn:fullDeri} can be expressed as 
\begin{equation}\label{eqn:chainRule4ExplicitSimp}
    \frac{\partial \psi_i}{\partial \rho_e} = 
    \delta_{ie}\frac{A}{A_i}\bm{\lambda}^T \frac{\partial^2 \bm{K}}{\partial \rho^2_e} \bm{U}
\end{equation}
where $\delta_{ie}$ denotes the Kronecker delta. Substituting these expressions into the total derivative (Eq.~\ref{eqn:fullDeri}) gives
\begin{equation} \label{eqn:fullDeriExpressEle}
    \frac{\mathrm{d}\psi_i}{\mathrm{d}\rho_e}
    =
    \delta_{ie}\frac{A}{A_i}\bm{\lambda}^T \frac{\partial^2 \bm{K}}{\partial \rho^2_e} \bm{U}
    -
    \left(\frac{A}{A_i}\frac{\partial \bm{K}}{\partial \rho_i}\bm{\lambda}\right)^T
    \bm{K}^{-1}
    \frac{\partial \bm{K}}{\partial \rho_e}\bm{U}
    -
    \left(\frac{A}{A_i}\frac{\partial \bm{K}}{\partial \rho_i}\bm{U}\right)^T
    \bm{K}^{-1}
    \frac{\partial \bm{K}}{\partial \rho_e}\bm{\lambda}
\end{equation}
Substituting this expression into the aggregation formula for $\frac{d\Psi}{d\rho_e}$ (Eq.~\ref{eqn:psi2psistar}), and again using the symmetry of ($\bm{K}$), the derivative of the hot spot measure can be written as
\begin{equation} \label{eqn:fullDeriExpress}
    \frac{\mathrm{d}\Psi}{\mathrm{d}\rho_e}
    =
    w_{e}\frac{A}{A_e}\bm{\lambda}^T \frac{\partial^2 \bm{K}}{\partial \rho^2_e} \bm{U}
    \underbrace{\underbrace{-\sum_i w_i\left(\frac{A}{A_i}\frac{\partial \bm{K}}{\partial \rho_i}\bm{\lambda}\right)^T}_{\bm{b}^T_{\lambda}}\bm{K}^{-1}}_{\bm{\xi}^T}
    \frac{\partial \bm{K}}{\partial \rho_e}\bm{U}
    \underbrace{\underbrace{-\sum_i w_i\left(\frac{A}{A_i}\frac{\partial \bm{K}}{\partial \rho_i}\bm{U}\right)^T}_{\bm{b}^T_{U}}\bm{K}^{-1}}_{\bm{\eta}^T}
    \frac{\partial \bm{K}}{\partial \rho_e}\bm{\lambda}
\end{equation}
With the auxiliary adjoints identified in Eq.~\ref{eqn:fullDeriExpress}, the compact form is
\begin{equation} \label{eqn:fullDeriExpressSimp}
    \frac{\mathrm{d}\Psi}{\mathrm{d}\rho_e}
    =
    w_{e}\frac{A}{A_e}\bm{\lambda}^T \frac{\partial^2 \bm{K}}{\partial \rho^2_e} \bm{U}
    +\bm{\xi}^T
    \frac{\partial \bm{K}}{\partial \rho_e}\bm{U}
    +
    \bm{\eta}^T
    \frac{\partial \bm{K}}{\partial \rho_e}\bm{\lambda}
\end{equation}

For compliance minimization, the nominal adjoint field satisfies $\bm{\lambda}=-\bm{U}$. Substituting this relation into Eq.~\ref{eqn:fullDeriExpressSimp} collapses the two auxiliary adjoint systems into a single one. After redefining the remaining auxiliary adjoint vector through
\begin{equation}
\bm{K}\bm{\eta}=\underbrace{\sum_i w_i\frac{A}{A_i}\frac{\partial \bm{K}}{\partial \rho_i}\bm{U}}_{\bm{b}},
\end{equation}
one obtains
\begin{equation}
\frac{\mathrm{d}\Psi}{\mathrm{d}\rho_e} = 
2\bm{\eta}^T \frac{\partial \bm{K}}{\partial \rho_e}\bm{U} - 
w_e\frac{A}{A_e} \bm{U}^T \frac{\partial^2 \bm{K}}{\partial \rho_e^2}\bm{U}
\end{equation}

\section{Relation between strain-energy density and von Mises stress}\label{apdx:stressMeasures}

For isotropic linear elasticity, the local stress-energy density ($W$) appearing in the compliance sensitivity can be written in terms of the principal stresses as
\begin{equation}\label{eq:strain_energy_principal}
2W = \bm{\sigma}^{T}\bm{D}_0^{-1}\bm{\sigma}
 = \frac{1}{E_0}\left[\sigma_1^2+\sigma_2^2+\sigma_3^2-2\nu\left(\sigma_1\sigma_2+\sigma_2\sigma_3+\sigma_3\sigma_1\right)\right],
\end{equation}
where $\sigma_1$, $\sigma_2$, and $\sigma_3$ are the principal stresses. The von Mises stress is given by
\begin{equation} \label{eq:vonmises_principal}
\sigma_{vM}^2
=\frac{1}{2}\left[(\sigma_1-\sigma_2)^2 + (\sigma_2-\sigma_3)^2+(\sigma_3-\sigma_1)^2\right].
\end{equation}
Equivalently, the strain-energy density may be decomposed as
\begin{equation} \label{eq:strain_energy_decomposition}
W = \frac{\sigma_{vM}^2}{6G_0} + \frac{\sigma_m^2}{2K_0},
\qquad 
\sigma_m=\frac{\sigma_1+\sigma_2+\sigma_3}{3},
\end{equation}
where
\begin{equation}
G_0=\frac{E_0}{2(1+\nu)},
\qquad
K_0=\frac{E_0}{3(1-2\nu)}
\end{equation}
are the shear and bulk moduli of the solid material, respectively. This decomposition shows that the strain-energy density contains both deviatoric and volumetric contributions, whereas the von Mises stress depends only on the deviatoric stress state. Consequently, similar spatial localization of the compliance sensitivity and von Mises stress may occur when the deviatoric contribution dominates, but no general one-to-one correspondence should be expected. For plane stress, Eq.~\ref{eq:strain_energy_principal} reduces directly by setting $\sigma_3=0$, while under plane strain the generally nonzero out-of-plane stress is retained.

\section{Evolution of the relative SHoSP correction}\label{apdx:rkappa}

\begin{figure}
    \centering
    \includegraphics[width=1.0\linewidth]{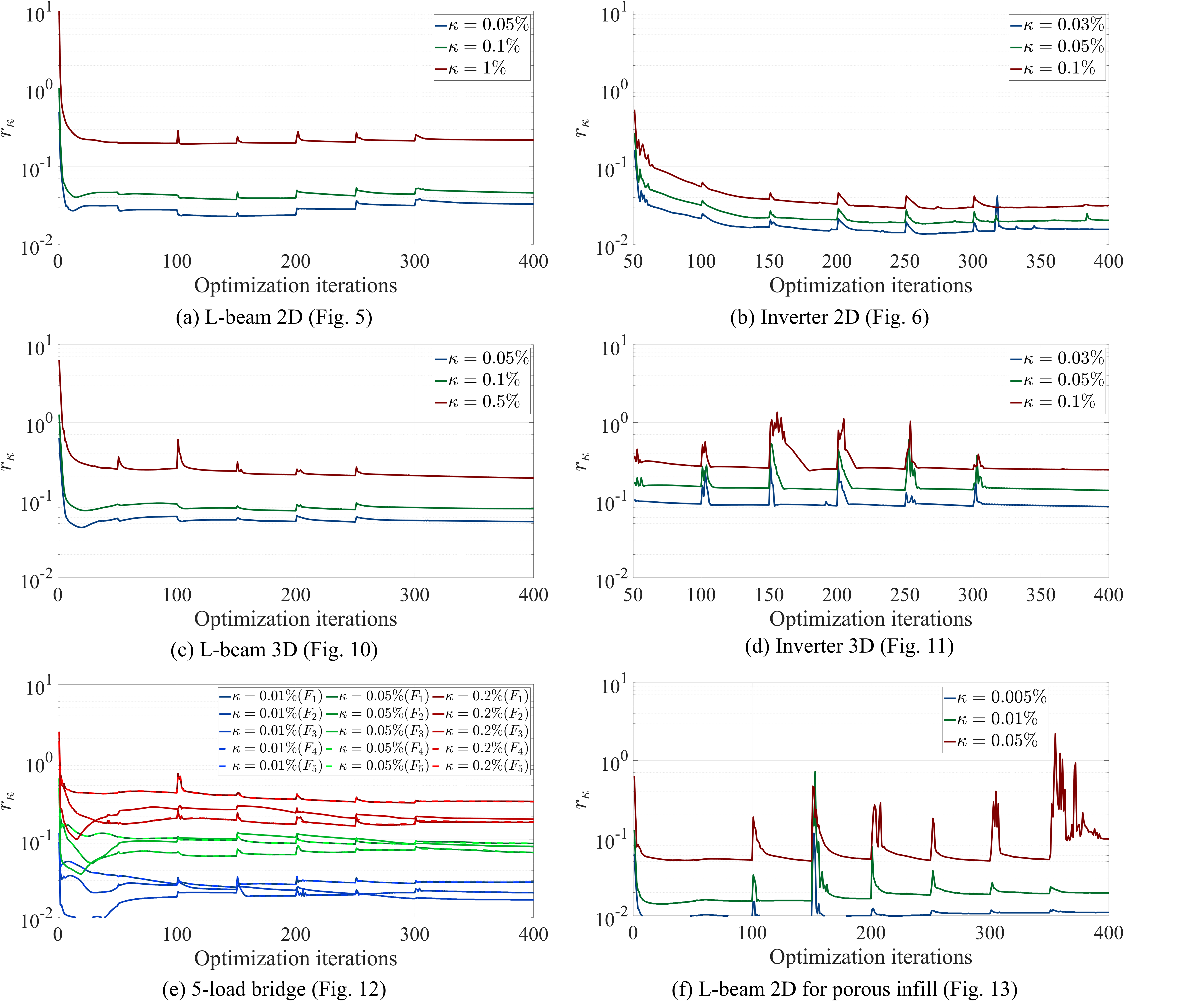}
    \caption{Evolution of the relative SHoSP correction $r_\kappa=\kappa\Psi/|\Phi|$ for the benchmark examples in Secs.~\ref{subsec:robustnessTest} and \ref{subsec:extendedFormulations}. Each subplot reports the SHoSP cases with nonzero $\kappa$ for the corresponding example. The histories provide numerical reference values for the penalization levels used in the paper and illustrate the transient increases that can occur after continuation updates, especially in compliant mechanism and porous infill problems.}
    \label{fig:rKappaStatistics}
\end{figure}

The relative SHoSP correction $r_\kappa=\kappa\Psi/|\Phi|$ was introduced in Sec.~\ref{subsec:penaltyStrength} as a diagnostic quantity for assessing the strength of the hot spot penalty relative to the nominal objective. Figure~\ref{fig:rKappaStatistics} reports the evolution of $r_\kappa$ for the benchmark examples in Sec.~\ref{sec:results} which provides numerical reference values for the penalization levels used in the controlled studies.

The histories show that moderate SHoSP settings generally lead to a non-negligible but not dominant correction relative to the nominal objective. They also reveal sharp transient increases after continuation updates in some cases, most visibly in compliant mechanism design and porous infill optimization. This behavior indicates that $r_\kappa$ can be useful not only as a post-processing quantity, but also as a practical monitor for detecting when the SHoSP correction temporarily becomes dominant during optimization.

\section{Convergence of the multi-load formulation}\label{apdx:multiConvergence}

\begin{figure}
    \centering
    \includegraphics[width=1.0\linewidth]{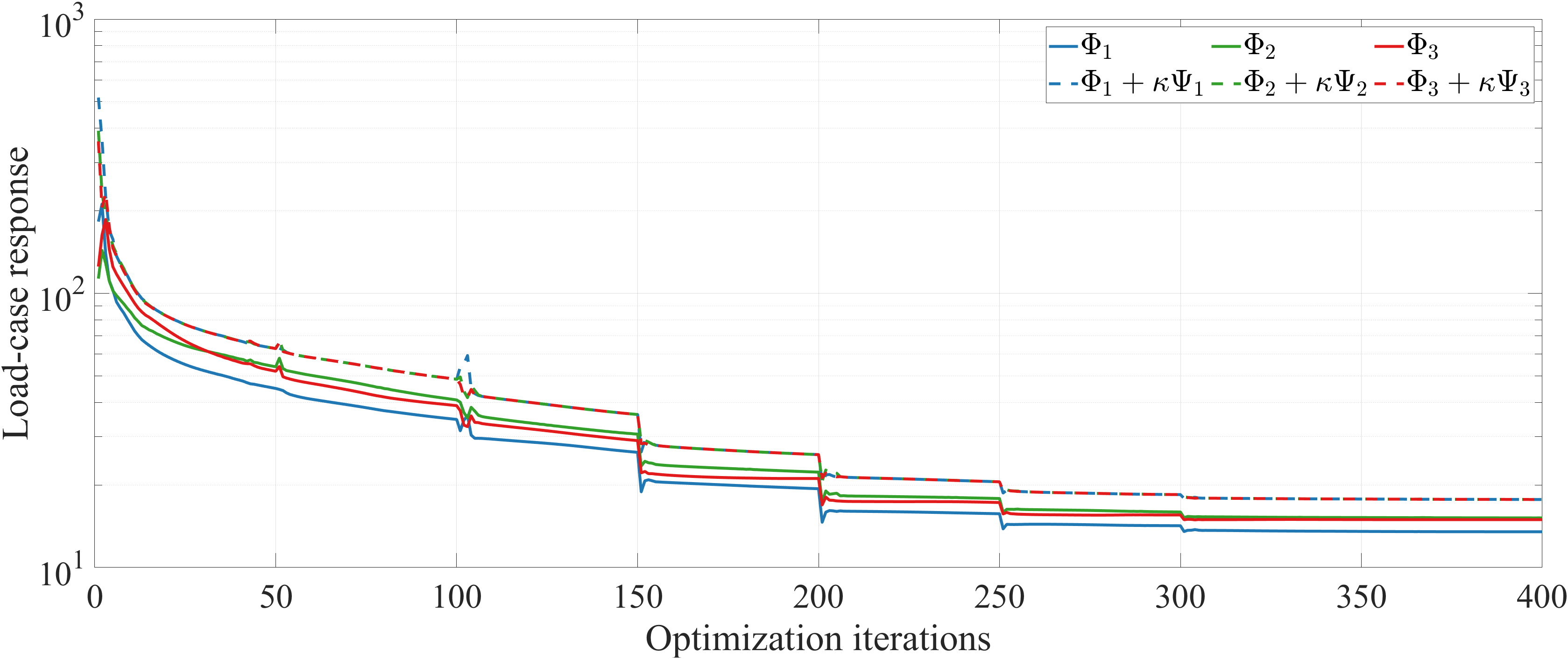}
    \caption{Convergence histories of the nominal load-case responses $\Phi_i$ and the corresponding augmented responses $\Phi_i+\kappa\Psi_i$ for the multi-load example with $\kappa=0.2\%$ in Fig.~\ref{fig:5loadBridge}. Owing to symmetry, only the three distinct load cases $F_1$--$F_3$ are shown.}
    \label{fig:5loadBridge_PhiPsi}
\end{figure}

Figure~\ref{fig:5loadBridge_PhiPsi} shows the convergence histories of the nominal load-case responses $\Phi_i$ and their SHoSP-augmented counterparts $\Phi_i+\kappa\Psi_i$ for the multi-load example considered in Fig.~\ref{fig:5loadBridge}, using the case with $\kappa=0.2\%$. The histories show that the SHoSP contribution remains active for all distinct loading conditions throughout the optimization, supporting the observation that sensitivity localization is suppressed across the full set of load cases rather than only in the governing response.

\bibliographystyle{plainurl} 
\bibliography{main}

\end{document}